\documentclass{elsart}

\usepackage{epsfig,amssymb,latexsym}
\journal{Annals of Physics}
\begin{document}
\begin{frontmatter}

 \newtheorem{theorem}{Theorem}
 \newtheorem{lemma}{Lemma}
  \newtheorem{remar}{Remark}
  \newtheorem{proposition}{Proposition}
  \newtheorem{example}{Example}

\title{Intertwining technique for the one-dimensional stationary
Dirac equation}

\author[Valladolid]{L.M. Nieto\corauthref{cor1}},
\corauth[cor1]{Corresponding author.}
\ead{luismi@metodos.fam.cie.uva.es}
\author[Tomsk]{A.A. Pecheritsin},
\author[Valladolid,Tomsk]{Boris F. Samsonov}

\address[Valladolid]{Departamento de F\'{\i}sica Te\'orica, Universidad de
Valladolid,  47005 Valladolid, Spain}

\address[Tomsk]{Department of Quantum Field Theory, Tomsk State University,
36 Lenin Ave., 634050 Tomsk, Russia}


\begin{abstract}
The technique of differential intertwining operators
(or Darboux transformation operators)
 is systematically
applied to the one-dimensional Dirac equation.
 The following aspects are investigated:
 factorization of a polynomial of Dirac Hamiltonians,
 quadratic supersymmetry, closed extension of
 transformation operators, chains of transformations,
 and finally
 particular cases of
 pseudoscalar and scalar potentials.  The method is widely
 illustrated by numerous examples.
\end{abstract}

\begin{keyword}
Dirac equation \sep  exact solutions   \sep  intertwining technique \sep  Darboux transformation
\PACS 02.30.Jr \sep 02.30.Tb \sep 03.65.-w \sep 11.30.Pb
\end{keyword}
\end{frontmatter}

\section{Introduction}

In recent times a growing interest to applications of
supersymmetric quantum mechanics
 (SUSY QM)  in different fields of theoretical and mathematical
 physics is noticed. Thus, it has been applied to generate new
 families of exactly solvable potentials
 \cite{FernHussMeln}--\cite{Samsonov99}, to study
ordinary symmetries of the set of Riccati equations \cite{CRF}, and
also to analyse
 a new kind of
 symmetry of ordinary differential equations, called
 translational invariance with respect to Darboux transformations
 \cite{FMRS}.
 As it has been understood after the work by Andrianov et al.
 \cite{Andr}, SUSY QM (first introduced by Witten
 \cite{Wit} as a toy model in Quantum Field Theory)
 is basically
 equivalent to a method of finding solutions of a differential
 equation from the known solutions of another equation just
 by applying to them a differential operator.
 This approach was studied for the first time by Darboux \cite{Darboux}
 and is known nowadays as the method of Darboux transformations.
Its most famous applications are related with nonlinear equations of
mathematical physics \cite{MatveevSall}
and inverse scattering problem \cite{Sukumar}--\cite{Bay}.
 Another formulation of the same method is due to
 Schr\"odinger and it is known as
 factorization method in quantum mechanics
  \cite{Schrodinger,InfeldHall}.

Though this method is widely used for
Schr\"odinger-like equations, its application to systems of
equations, and in particular to the Dirac system,  is studied much
less. Probably the first paper where a differential-matrix
intertwining operator for the Dirac equation is constructed is due
to Anderson \cite{Anderson}. It deals with component equations for
the transformation operator. For the particular case of a
pseudoscalar potential this author has found rather complicated
equations defining the transformation operator. This might be the
reason why his method has not received much popularity. In another
paper \cite{Stahlhofen} Stahlhofen applies a general
theorem proved in  \cite{MatveevSall} to study transparent
pseudoscalar potentials for the Dirac equation,
obtaining and analysing new kind of potentials with eigenvalues
embedded in the continuous spectrum.
Although there is not a new technique in this paper,
it illustrates the idea that similar methods may find unexpected
applications.
We notice also the papers \cite{Sall-82,Sall-87}
(for a review see  \cite{MatveevSall}) where {\sl ad hoc}
Darboux transformation operators are constructed for
some systems of differential equations,
and the paper \cite{DascalovChristov} where the authors
use differential transformation operators for finding particular
solutions of the inverse problem for a regular Dirac operator.

We would like to remark that the technique of integral intertwining
 operators  is well-known in the method of inverse  quantum
scattering \cite{PratsToll} (see also \cite{LevitanSargsyan}), but
differential-matrix intertwiners have been found only quiet recently
\cite{DascalovChristov,SamsonovPecheritsin}.

In this paper
we give a consistent introduction to
differential-matrix intertwining operators for the one-dimensional
stationary Dirac equation (Section 2) and study their main
properties. We shall show that every property of a differential
intertwiner for the Schr\"odinger equation finds its counterpart
in the case of the Dirac equation. Thus, in Section 3 we prove a
theorem about factorization of a polynomial of Dirac operators by
means of
Darboux transformation operators, which serves as a basis for
establishing a hidden quadratic supersymmetry underlying our
method (Section 6). In Section 4 we show that despite the fact
that our transformation operators have nontrivial kernels, a
one-to-one correspondence between the spaces of solutions of the
initial and transformed equations is possible.  Section 5 is
devoted to analysing  transformation operators as acting in a
Hilbert space. In Section 7 we study chains of transformations and
establish relativistic analogs of Crum determinant formulas \cite{Crum}.
Sections 8 and 9 are devoted to studying particular cases of
pseudoscalar and scalar potentials. Our method is widely
illustrated by numerous examples given in Section 10. Finally, in
Section 11 we draw some conclusions and give an outlook
of future works.

\section{Darboux transformation for the Dirac equation}

Consider the one-dimensional stationary Dirac equation
\begin{equation}  \label{dirac0}
h_0 \psi = E \psi\,.
\end{equation}
Here $h_0 = \gamma \partial_x + V_0$, $\gamma = i \sigma_2$,
$\sigma_2$ is a Pauli matrix, $\partial_x\equiv d/dx$,
 $V_0$ is $2\times 2$ real
symmetric potential matrix and the function $\psi $  is a
two-component vector (we shall call it spinor), $\psi = (\psi_1,
\psi_2)^t$ (the superscript ``$t$" meaning the transpose).
We do not indicate the interval where
the real variable $x$ falls, because our constructions are independent of
this interval.

Suppose we know all solutions $\psi$ to Eq.~(\ref{dirac0})
and we want to solve a similar equation for
another Hamiltonian $h_1$ defined by the potential $V_1$. The
problem of finding eigenfunctions $\varphi $ of the Hamiltonian $h_1$,
\begin{equation} \label{dirac1}
h_1 \varphi = E \varphi,
\end{equation}
can be reduced to the problem of finding  a
{\it transformation operator}  (see e.g. \cite{LevitanSargsyan}),
also called {\it intertwiner} \cite{CRF},
defined by the equation
\begin{equation} \label{splet}
Lh_0 = h_1L   \,.
\end{equation}
 It follows from here that the function
 $\varphi =  L\psi$ is a solution (maybe trivial)
  of Eq.~(\ref{dirac1}).
  Here and in the following we shall suppose that all operators act
 in the space of spinors with sufficiently smooth
 (i.e. infinitely differentiable)  components.
 In this paper we restrict ourselves to differential
 transformation operators.
 From this point of view let us consider the simplest
 {\it differential}
 transformation operator
\begin{equation}
 L = A \partial_x + B\,.
 \end{equation}
Here $A$ and $B$ are $2\times 2$ matrices with $x$-dependent entries.
 Similar transformation operator for the Schr\"odinger equation leads
 (see e.g. \cite{BSTMF}) to the
 transformation investigated for the first time by Darboux
 \cite{Darboux}), and therefore it is usually called ``Darboux
 transformation operator". In this context it is natural to call
 any differential transformation operator Darboux transformation
 operator \cite{BSTMF}.

 Using the intertwining relation (\ref{splet}), and  assuming the
 linear independence of derivative operators $\partial _x^k$,
 $k=0,1,\ldots $,
 we
 are led to the following system of equations for $A$, $B$ and $V_1$:
 \begin{eqnarray}
& & A\gamma - \gamma A = 0\,, \label{eqa}\\ [1ex]
& & A V_0 - V_1 A + B \gamma - \gamma B - \gamma A_x = 0\,, \label{eqv}\\
[1ex] & & A V_{0x} + B V_0 - \gamma B_x - V_1 B = 0\,. \label{eqb}
 \end{eqnarray}
 The subscript $x$ denotes the derivative with respect to $x$,
 e.g. $A_x = dA/dx$, and the matrix $A_x$ is composed of
 the first derivatives of the elements of the matrix $A$.

 The Eq.(\ref{eqa})
 may be considered as a restriction imposed on the matrix $A$:
 two of its four elements may be chosen
 arbitrarily, and the other two should be found in terms of the previous
 ones.
 The Eq. (\ref{eqv}) defines the potential
 difference $D = V_1 - V_0$:
       \begin{equation} \label{eqd}
   D = (A V_0 - V_0 A + B \gamma - \gamma B - \gamma A_x) A^{-1}\,.
       \end{equation}
 We are supposing that $A$ is an invertible matrix for all values of $x$.
 From (\ref{eqb}) we will find $B$. To integrate this equation we
 first change the dependent variable $B$ in favor of a new
 variable $\Omega$: $B=A\Omega$. Thus, upon using
 (\ref{eqd}) we obtain from (\ref{eqb}) the equation for
 $\Omega$
 \begin{equation}
 (V_0 - \gamma \Omega)_x + \Omega V_0 - V_0 \Omega +
 (\gamma \Omega -  \Omega \gamma) \Omega = 0.
 \label{eqsig}
 \end{equation}
 This equation can be considered as
 the derivative of
  a relativistic analog of the Riccati
 equation
 that appears when
 similar calculations are carried out on the Schr\"odinger equation.
 Eq.~(\ref{eqsig}) can be linearized and integrated by a similar
 substitution: $\Omega = - {\mathcal U}_x\, {\mathcal U}^{-1}$.
 After such a calculation we get
  \begin{equation}
 [\,{\mathcal U}^{-1}(V_0{\mathcal U} + \gamma {\mathcal U}_x)\,]_x  = 0  \,,
 \label{equ1}
 \end{equation}
  where ${\mathcal U}$ is supposed to be everywhere invertible.
Now, denoting by $\Lambda $ a matrix integration constant,
we get the equation for ${\mathcal U}$:
 \begin{equation}    \label{equ2}
 \gamma {\mathcal U}_x + V_0{\mathcal U}  = {\mathcal U} \Lambda \, .
  \end{equation}
 The matrix $\Lambda $ should be taken hermitian.

 We notice  that (\ref{equ2}) is very similar to
 the initial Dirac Eq.~(\ref{dirac0}).
 The single difference is that
 $\psi $ is a spinor and $E$ is a number, while ${\mathcal U}$ and $\Lambda$ are
 matrices.
 Nevertheless,
 if solutions of (\ref{dirac0}) are known we can immediately solve (\ref{equ2}).

 Since the matrix $\Lambda $ is hermitian it always can be reduced
 to a diagonal form.
 Therefore let us take $\Lambda $ to be diagonal
\begin{equation} \label{diagLambda}
\Lambda =\left(
\begin{array}{cc}
\lambda _1 & 0\\
0& \lambda _2
\end{array}
\right).
\end{equation}
  If  we now choose
 ${\mathcal U} = (u_1,u_2)$, where spinors  $u_1$ and $u_2$ are solutions of
 (\ref{dirac0}) with the eigenvalues $\lambda _1$ and $\lambda _2$
 respectively,
$$
    h_0u_{1,2}=\lambda_{1,2}u_{1,2}\,,
$$
then ${\mathcal U}$ is a solution of (\ref{equ2}). We will
 choose the spinors $u_1$ and $u_2$ real, which is always possible
 if we suppose $\lambda _{1}$ and $\lambda _{2}$ to be real.

 Once the matrix ${\mathcal U}$ is known,
 the transformation operator $L$ and the transformed potential
 $V_1$ are defined up to the matrix $A$
 $$
 L = A[\partial_x - {\mathcal U}_x\, {\mathcal U}^{-1}]\,,
 $$
$$
V_1 = A(V_0 + \Omega \gamma - \gamma \Omega -  \gamma
 A^{-1} A_x) A^{-1}\,.
$$
 The matrix $A$ remains arbitrary provided it satisfies the
 condition (\ref{eqa}). Nevertheless, remark that the simple gauge
 transformation $\varphi =A\widetilde\varphi $ reduces it to the
 identity.
 Therefore, without
 loss of  generality
 we can take $A=I$, where $I$ is $2\times 2$ unit matrix. Now the
operator $L$ has its
 simplest possible form
 \begin{equation} \label{eql}
 L = \partial_x - {\mathcal U}_x\, {\mathcal U}^{-1} \,,
 \end{equation}
  which is the relativistic extension of the well known Darboux
  transformation operator for the Schr\"odinger equation (see e.g.
  \cite{BSTMF}).
 For the transformed potential we get
 \begin{equation}\label{V1}
 V_1 = V_0 + D, \quad D = \Omega \gamma - \gamma \Omega,
  \quad \Omega = - {\mathcal U}_x\, {\mathcal U}^{-1}\, .
 \end{equation}
  We observe here that the transformation operator and
  the transformed potential are completely defined by the matrix-function
  ${\mathcal U}$. Therefore, we will call it {\it transformation function}.

 Using Eq.~(\ref{equ2}) one can express ${\mathcal U}_x$
 in terms of ${\mathcal U}$ and $V_0$. This gives us another representation
 for the transformed potential
  \begin{eqnarray}  \label{V1D}
  V_1  &  =  &  - \gamma V_0 \gamma + \widetilde D,\\ [1ex]
 \widetilde{D}  & = &  u \Lambda u^{-1} + \gamma u \Lambda
u^{-1}\gamma = (\lambda_1 - \lambda_2) \Delta \left(\begin{array}{cc}
d_1 & d_2 \\ d_2 & -d_1
\end{array}\right) \,,
\label{d-matr}
\end{eqnarray}
where
\begin{equation}\label{d12}
 d_1 = u_{11}u_{22} + u_{12}u_{21}\,,\qquad
 d_2 = u_{21}u_{22} + u_{11}u_{12} \,,
\end{equation}
$\Delta = (\det {\mathcal U} )^{-1}$ and $u_{ij}$, $i,j=1,2$, are the entries of
the matrix
${\mathcal U}$. Similar transformations
applied to the formula (\ref{eql}) yield
\begin{equation}\label{Lsimpl}
 \varphi =L\psi =\gamma ({\mathcal U}\Lambda {\mathcal U}^{-1}-E)\psi  \,.
\end{equation}
 This formula is equivalent to  (\ref{eql}) provided $L$ acts on
 an eigenfunction of $h_0$ with the eigenvalue $E$.

Before going further, it is necessary to point out a couple of simple but
relevant comments.
\begin{remar}\label{rem0}
Given a  hamiltonian $h_{0}=\gamma \partial_x + V_{0}$, if the
hermitian potential matrix
$V_0(x)$ is symmetric, then it can be reduced by a smooth orthogonal
transformation {\rm \cite{LevitanSargsyan}} to the form
 \begin{equation}\label{CanonV}
V_0(x)=p_0(x)\sigma_3+q_0(x)\sigma_1,
 \end{equation}
where $\sigma_1,\sigma_3$ are Pauli matrices, known as a {\sl canonical
representation} of $V_0$.
Therefore, in the following we
shall use only this canonical representation for the Dirac potentials.
\end{remar}

\begin{remar}\label{rem1}
Any potential of the form {\rm (\ref{CanonV})} has the properties
\begin{equation}
\label{gVg}
\gamma V_0\gamma =V_0\,, \qquad \gamma V_0+V_0\gamma =0\,.
\end{equation}
\end{remar}

 Note that the operator $L$ has a nontrivial kernel, $L\,{\mathcal U} =0$, i.e.
 $\mbox{ker} L$ is a two dimensional linear space spanned by the
 spinors $u_1$ and $u_2$. This means that the set of the spinors
 $L\psi $, when $\psi $ runs through the whole space of solutions
 of the initial equation, may
 in general
 not coincide with the whole space of
 the solutions of the transformed equation.
 In Section~3 we shall show that despite this fact a one-to-one
 correspondence between the spaces of solutions of the equations
 (\ref{dirac0}) and (\ref{dirac1}) may be established.

\section{One-to-one correspondence between the spaces of solutions}

To define the Dirac Hamiltonian and the transformation operator
 we have not used the notion of the Hilbert space
 and an inner product is not defined in the space of spinors.
 Nevertheless, we need operators adjoint to given ones.
 Therefore in the space of operators
 we are working we shall
 define the operation of conjugation in a formal way.
 Namely, we will require
 $\partial_x^+=-\partial_x$, $(AB)^+=B^+A^+$,
 $i^+=-i$ (where $i$ is the imaginary unity),
 and if $A$ is a
 matrix, $A^+$ is its hermitian conjugate in the usual sense
 (i.e., it is obtained from $A$ by transposing and taking the complex
 conjugation). Moreover, for a given invertible matrix ${\mathcal U}$ we will
assume that the operations of taking the inverse and adjoint commute, i.e.,
 \begin{equation}\label{upm}
 ({\mathcal U}^+)^{-1}=({\mathcal U}^{-1})^+    \,.
 \end{equation}

 It is easy to see that the Dirac operators $h_{0}$ and $h_{1}$ are
self-adjoint. Therefore, the adjoint intertwining relation reads
\begin{equation} \label{spletsopr}
h_0L^+ = L^+h_1 \,.
\end{equation}
This means that the operator $L^+$ is also a transformation
operator and it realizes the transformation in the opposite
direction, from the solutions of (\ref{dirac1}) to solutions of
(\ref{dirac0}). But as we will show below, it is not an inverse of $L$.

It is well known that for any fixed value of the energy the Dirac system
 (\ref{dirac0}) has two linearly independent solutions.
As in the case of the Schr\"odinger equation,
 if we know one of them, $\psi$ with  eigenvalue
 $E$, the other solution $\widetilde \psi$ with the same eigenvalue can be found by a
 quadrature.
 In this way we get the spinors $\widetilde u_{1}$
 and $\widetilde u_{2}$ with
 eigenvalues     $\lambda_{1}$ and $\lambda_{2}$.
 They  do not belong to the space $\mbox{ker}L$ and hence $v_{1,2}=L\widetilde
u_{1,2}$
 are
 nontrivial solutions of the transformed equation with the
 eigenvalues $\lambda_{1,2}$.  We will find them next.

 For this purpose we need a counterpart of the Wronskian
 which, like in the case of the Schr\"odinger equation, may be defined as  a function
 of two solutions $\psi $ and $\widetilde\psi$ corresponding to the same eigenvalue
$E$, which is independent of the variable $x$. Using the adjoint form of the
Eq.~(\ref{dirac0}), we easily find
  \begin{equation}
   (\widetilde\psi^+ \gamma \psi)_x =
   \widetilde\psi_{1}\psi_{2x}+\widetilde\psi_{1x}\psi_{2} -
   \widetilde\psi_{2}\psi_{1x}-\widetilde\psi_{2x}\psi_{1}=0\,,
    \end{equation}
   which means that
  \begin{equation} \label{W}
   W(\widetilde\psi ,\psi ):=\widetilde\psi^+ \gamma \psi
   =\widetilde\psi_1\psi_2-\widetilde\psi_2\psi_1=
   \mbox{constant} \,.
    \end{equation}
  Hence, the function $W(\widetilde \psi ,\psi )$ we just defined can  play the role of
the Wronskian.
The solutions $\psi $ and $\widetilde\psi $  can always be chosen such that
the constant that appears in (\ref{W})  is equal to one.

 The Dirac system with a canonical potential $V_0=p_0\sigma_3+q_0\sigma_1$,
 written in terms of the components $\psi_1$ and $\psi_2$ of the spinor
$\psi=(\psi_1,\psi_2)^t$, has the form:
 \begin{equation}
 \label{u12}
    \psi_{2x}+p_0\psi_1+q_0\psi_2 = E\psi_1 \,, \qquad
 - \psi_{1x}+q_0\psi_1-p_0\psi_2 = E\psi_2 \,.
 \end{equation}
The same system for the components of the spinor $\widetilde\psi
=(\widetilde\psi_1,\widetilde\psi_2)^t$ is
 \begin{equation}
    \widetilde\psi_{2x}+p_0\widetilde\psi_1+
q_0\widetilde\psi_2 = E\widetilde\psi_1 \,,
\qquad
 - \widetilde\psi_{1x}+q_0\widetilde\psi_1-p_0\widetilde\psi_2 = E\widetilde\psi_2 \,.
 \label{u12t}
 \end{equation}
  If now we eliminate the function $q_0$ from the last equations of  (\ref{u12})
and (\ref{u12t}) we obtain
  \begin{equation}
 \psi_{1}\widetilde\psi_{1x}-  \widetilde\psi_{1}\psi_{1x} =p_0+E\,,
  \end{equation}
  where we have used formula (\ref{W}).
  From here, assuming that $\psi_1$ is not identically zero,  we easily find
    \begin{equation} \label{wtpsi1}
  \widetilde\psi_1=\psi_1\int_{x_0}^x\frac{p_0+E}{\psi_1^2}\,dx  \,.
    \end{equation}
    Formula (\ref{W}) gives us the second component of
    the spinor $\widetilde\psi$:
     \begin{equation}   \label{wtpsi2}
     \widetilde\psi_2=\frac{\widetilde\psi_1\psi_2}{\psi_1}-\frac{1}{\psi_1} \,.
     \end{equation}
If $\psi_1(x)\equiv 0$ then, for a
nontrivial spinor $\psi=(\psi_1,\psi_2)^t$,
$\psi_2(x)$
is not identical null, and the following alternative formulae should be used:
    \begin{equation}
  \widetilde\psi_2=\psi_2\int_{x_0}^x\frac{E-p_0}{\psi_2^2}\,dx\,, \qquad
\label{wtpsi4}
     \widetilde\psi_1=\frac{\widetilde\psi_2\psi_1}{\psi_2}+\frac{1}{\psi_2}\,.
     \end{equation}

  Now, we can apply  (\ref{Lsimpl}) to the eigenspinors
  $\widetilde u_{1,2}$ with the eigenvalues $\lambda_{1,2}$ which gives
  us the action of $L$ on the matrix $\widetilde{\mathcal U}=(\widetilde
u_1,\widetilde u_2)$.
  After some simple algebra we get
  \begin{equation}
\label{qwe}
  L\,\widetilde {\mathcal U}=(\lambda_2-\lambda_1){\mathcal V}  \,,
  \end{equation}
  \begin{equation}\label{usm1}
  {\mathcal V} =({\mathcal U}^+)^{-1}                 \,.
  \end{equation}
  Hence, the matrix ${\mathcal V}$ satisfies the equation $h_1{\mathcal V} =
{\mathcal V}\Lambda$,
  $\Lambda=\mbox{diag}(\lambda_1 ,\lambda_2)$.
  It is composed of the spinors $v_1$ and $v_2$, ${\mathcal V} =(v_1,v_2)$,
  satisfying the Dirac equation with the potential $V_1$,
  $h_1v_{1,2}=\lambda_{1,2}v_{1,2}$.
  Two other eigenspinors of $h_1$, $\widetilde v_1$ and $\widetilde v_2$,
  with respective eigenvalues $\lambda_1 $ and $\lambda_2$,
  $h_1\widetilde v_{1,2}=\lambda_{1,2}\widetilde v_{1,2}$, may be found with
  the help of the same formulas (\ref{wtpsi1})--(\ref{wtpsi4}),
  applied this time to the transformed equation.

     Hence,  we have established a one-to-one correspondence
     between the spaces of solutions of the equations
     (\ref{dirac0}) and (\ref{dirac1}).
     For any $E\ne \lambda_{1},\lambda_{2}$ the operators $L$ and $L^+$
     realize this correspondence; if $E=\lambda_{1,2}$  the correspondence can be
assured by the mapping ${\mathcal U}_{1,2}\leftrightarrow \widetilde {\mathcal
V}_{1,2}$,
$\widetilde
{\mathcal U}_{1,2}\leftrightarrow {\mathcal V}_{1,2}$,  considered as a linear mapping.

As a final remark of this Section, we notice that $L^+{\mathcal V}=0$, meaning that
$\widetilde{\mathcal U} \in {\rm ker} L^+L$. In addition, the operator $L^+$
expressed in terms of ${\mathcal V}$ reads as
$L^+=-\partial_x-({\mathcal U}^+)^{-1}{\mathcal U}^+_x=-\partial_x+{\mathcal V}_x{\mathcal
V}^{-1}$. This means that if we interchange the role of the initial and final
equations, the function ${\mathcal V}$ becomes the transformation function for the
transformation operator of  the type (\ref{eql}), taken with the opposite sign
and realizing the transformation in the opposite direction, a fact that we
already mentioned. Moreover, for the transformed equation the function
$\widetilde {\mathcal V}$ plays the same role that $\widetilde{\mathcal U}$ plays for
the initial equation. This implies that, according to (\ref{qwe}),
$L^+{\mathcal V}=(\lambda_1-\lambda_2){\mathcal U}$. Taking into consideration the fact
that ${\mathcal U}\in {\rm ker}L$, we find that in one hand
$\widetilde{\mathcal V}\in {\rm ker}LL^+$. On the other hand, \
$\{{\mathcal U},\widetilde{\mathcal U}\} \in {\rm ker}(h_0-\lambda_1)(h_0-\lambda_2)$,
and $\{{\mathcal V},\widetilde{\mathcal V} \}\in {\rm
ker}(h_1-\lambda_1)(h_1-\lambda_2)$. Hence, we see that
 \begin{equation}\label{kerrr1}
{\rm ker}(h_0-\lambda_1)(h_0-\lambda_2)={\rm ker} L^+L={\rm span}
({\mathcal U},\widetilde{\mathcal U})
 \end{equation}
and
 \begin{equation}\label{kerrr2}
{\rm ker}(h_1-\lambda_1)(h_1-\lambda_2)={\rm ker} LL^+={\rm span}
({\mathcal V},\widetilde{\mathcal V})\,.
 \end{equation}
From these properties, we can suspect that the operator $L^+L$ coincides with
$(h_0-\lambda_1)(h_0-\lambda_2)$ and the operator $LL^+$ coincides with
$(h_1-\lambda_1)(h_1-\lambda_2)$. These facts will be proved in the next Section.

\section{Factorization property of Darboux transformations}

Before establishing the main result of this Section, we need first to prove some
auxiliary relations.

\begin{proposition}\label{lm1}
For any real $2\times 2$ matrix $A$ the following
formula holds: $\gamma A+A^+\gamma ={\rm tr}A\,\gamma$.
\end{proposition}

{\it Proof.}
Since the set of matrices
$\left\{I,\gamma ,\sigma_1,\sigma_3\right\}$
 is complete in the space of $2\times 2$ matrices, any
such a matrix may be presented in the form
$$
A=aI+b_1\sigma_1+b_2\gamma+b_3\sigma_3
$$
and  therefore $\mbox{tr}A=2a$. The statement of the Proposition follows
from the fact that the Pauli matrices  anticommute.
\hfill$\Box$

\begin{lemma}\label{lm2}
 The matrix ${\mathcal U}$ defined as
 ${\mathcal U} :=(u_1,u_2)$,
 where $u_1$ and $u_2$ are real eigenspinors of $h_0$ (with eigenvalues
 $\lambda_1$ and $\lambda_2$ respectively) chosen such that
 ${\mathcal U}$ is non-degenerate,
 satisfies the following relationship:
\begin{equation} \label{U}
U:={\mathcal U}_x\, {\mathcal U}^{-1}-({\mathcal U}_x\, {\mathcal U}^{-1})^+
=-(\lambda_1+\lambda_2)\gamma
\,.
\end{equation}
\end{lemma}

{\it Proof.}
 The equation for ${\mathcal U}$ (\ref{equ2}) together with   (\ref{upm})
 implies
 \begin{equation}
({\mathcal U}^+)^{-1}{\mathcal U}_x^+=-V_0\gamma +({\mathcal U}^+)^{-1}\Lambda  {\mathcal
U}^+\gamma
\,.
 \end{equation}
From here it follows the relation for $U$ as given in (\ref{U})
\begin{equation}
U=\gamma V_0+V_0\gamma -\gamma {\mathcal U}\Lambda {\mathcal U}^{-1}- ({\mathcal U}\Lambda
{\mathcal U}^{-1})^+\gamma \,,
 \end{equation}
  which, after taking into account Remark \ref{rem1},
  can be rewritten in the equivalent form
\begin{equation}
U=-[\gamma {\mathcal U}\Lambda {\mathcal U}^{-1}-({\mathcal U}\Lambda {\mathcal
U}^{-1})^+\gamma ]\,.
\end{equation}
The statement of the Lemma \ref{lm2}
 follows from this equation, Proposition
\ref{lm1}, and the following equality:
 $\rm{tr}\, ({\mathcal U}\Lambda {\mathcal U}^{-1})=\mbox{tr}\,\Lambda
 =\lambda_1+\lambda_2$.
\hfill$\Box$

\begin{proposition}
The following relationships hold for the matrices ${\mathcal U}$, chosen as in
Lemma~\ref{lm2}, and $\Lambda$, given in {\rm (\ref{diagLambda})}:
 \begin{eqnarray} \label{rel1}
  {\mathcal U}_{xx}{\mathcal U}^{-1}=V_0^2+\gamma V_{0x}-{\mathcal U}\Lambda^2{\mathcal
U}^{-1} \,,
\\ [2ex]
\label{rel2}
 \Lambda^2-(\lambda_1+\lambda_2)\Lambda =-\lambda_1\lambda_2I\,. \quad
 \end{eqnarray}
\end{proposition}

{\it Proof.}
  The first one is a differential implication of the Eq.~(\ref{equ2}). Indeed, if we take
the derivative of this equation and
  replace the first derivative of ${\mathcal U}$ using the same Eq.~(\ref{equ2}),
we get
 \begin{equation}\label{z1}
  \gamma {\mathcal U}_{xx}+V_{0x}{\mathcal U} +V_0\gamma V_0{\mathcal U} -V_0\gamma
{\mathcal U}\Lambda  =
  \gamma V_0{\mathcal U}\Lambda -\gamma {\mathcal U}\Lambda^2  \,.
 \end{equation}
   Now we use the formulae
   \begin{equation}
    V_0\gamma V_0{\mathcal U} =-\gamma V_0^2{\mathcal U} \,,\qquad
    V_0\gamma {\mathcal U} \lambda =-\gamma V_0{\mathcal U}\Lambda  \,,
   \end{equation}
 which are direct implications of (\ref{gVg}), and rewrite
 (\ref{z1}) as follows:
 \begin{equation}
\gamma {\mathcal U}_{xx}+V_{0x}{\mathcal U} -\gamma V_0^2{\mathcal U} = -\gamma
{\mathcal U}\Lambda^2 .
  \end{equation}
  Formula (\ref{rel1}) is now evident.

\noindent
Formula (\ref{rel2}) may be proved straightforwardly, by computing
explicitly its left hand side using (\ref{diagLambda}).
\hfill$\Box$

We establish and prove now the main result of this Section.

\begin{theorem}  \label{th1}
Darboux transformation operator $L$ given in {\rm (\ref{eql})} and its
formally adjoint $L^+$, constructed with the help of the matrix ${\mathcal U}
:=(u_1,u_2)$,
 where $u_1$ and $u_2$ are real eigenspinors of $h_0$ (with eigenvalues
 $\lambda_1$ and $\lambda_2$ respectively) chosen such that
 ${\mathcal U}$ is non-degenerate, factorize the following polynomial of the Dirac
Hamiltonians
 $h_0$ and $h_1$:
\begin{eqnarray} \label{LPL}
L^+L & = &  (h_0-\lambda_1I)(h_0-\lambda_2I)\,,
\\ [1ex]
 \label{LLp}
LL^+  &  =  &  (h_1-\lambda_1I)(h_1-\lambda_2I)\,.
\end{eqnarray}
\end{theorem}

{\it Proof.}
  Let us consider the superposition $L^+L$. Using (\ref{eql}) and after
some simple algebraic
  rearrangements, one gets
   \begin{equation}
  L^+L=-\partial_x^2+U\partial_x+{\mathcal U}_{xx}\, {\mathcal U}^{-1}-U{\mathcal U}_x\,
{\mathcal U}^{-1} .
   \end{equation}
  We now replace here $U$ by the right hand side of the formula
  (\ref{U}) and use Eq.~(\ref{rel1}) obtaining
   \begin{equation}
   \begin{array}{rcl}
  L^+L=& -&\partial_x^2+V_0^2+\gamma V_{0x}-{\mathcal U}\Lambda ^2{\mathcal U}^{-1}\\
[1ex]
&-&   (\lambda_1+\lambda_2)(\gamma \partial_x+V_0)
 -{\mathcal U} [\Lambda^2- (\lambda_1+\lambda_2)\Lambda]{\mathcal U}^{-1}  \,.
    \end{array}
   \end{equation}
   Formula (\ref{LPL}) follows from here,  Eq.~(\ref{rel2}),
   and the equalities
   $\gamma \partial_x+V_0=h_0$ and
   $-\partial_x^2+V_0^2+\gamma V_{0x}=h_0^2$.

 To prove formula (\ref{LLp}), we act with $L$ on the Eq.~(\ref{LPL}), and take into
account the intertwining relation
 (\ref{splet}). As a result, we obtain
  \begin{equation} \label{LLP}
  LL^+L\psi =(h_1-\lambda_1I)(h_1-\lambda_2I)L\psi \,.
  \end{equation}
 The last equation means that (\ref{LLp}) is valid
 for any $\varphi =L\psi$ and it remains to prove it for the spinors
 which can not be represented in this form.
 But we know that such
 spinors are eigenspinors of $h_1$, either with the eigenvalue
$\lambda_1$
  or with $\lambda_2$. For these spinors both right and left
  hand sides of (\ref{LLp}) are zero, since the spaces
  $\mbox{ker}\,LL^+$ and $\mbox{ker}\,(h_1-\lambda_1I)(h_1-\lambda_2I)$
  coincide, as we already commented in (\ref{kerrr2}).
\hfill$\Box$

As a consequence of this Theorem, taking into account (\ref{d-matr})
and Remark 2, summarized in Eq.~(\ref{gVg}),
the Darboux transformations
keep unchanged the canonical representation of a potential.

To conclude this Section, we notice that for the case
$\lambda_1=-\lambda_2=\lambda_0$ Eq.~(\ref{LPL}) takes the form
$$
L^{+}L = (h_0 - \lambda_0)(h_0 + \lambda_0) = (h_0^2 - \lambda_0^2),
$$
a result that was previously reported in \cite{Anderson} for a
scalar potential.

 \section{Darboux transformation operators in a  Hilbert space}

 Let us consider a Dirac Hamiltonian $h_0$ defined in a dense domain
 $D$ of the Hilbert space
 $H=L^2({\mathbb R})\otimes {\mathbb  C}^2$, $D\subset H$
 with the inner product
 \begin{equation} \label{IP}
 \langle  \Psi |\Phi \rangle  :=
 \int _{-\infty}^{\infty}\psi^\dagger (x)\varphi (x)dx \,.
 \end{equation}
 Here $\psi^\dagger (x)=(\psi_1^*(x),\psi_2^*(x))$.
 We shall denote the elements of the Hilbert space by capital
 symbols and we shall keep small symbols for the
 coordinate representation of the corresponding spinors.
 We shall suppose that the differential expression
 $\gamma \partial_x + V_0$, initially defined in a dense subset
 $D\subset L^2({\mathbb R})\otimes {\mathbb  C}^2$
 of  sufficiently smooth functions, has a self-adjoint extension with domain of
   definition $\overline{D} $,
 which we shall denote by the same symbol $h_0$. In this case
 the system of eigenspinors $|\Psi_E\rangle  \in D$,
 $h_0|\Psi_E\rangle =E|\Psi_E\rangle $,  is complete in $H$.
 This means that there exists a  measure $\rho (E)$ such that the
 following completeness condition takes place
\begin{equation}
 \int d\rho(E)\,|\Psi_E\rangle \langle \Psi_E|=1   \,.
\end{equation}
 This equation should be understood in a week sense, i.e., it is equivalent to
 \begin{equation}
   \int d\rho(E)\,\langle 
   \Phi_n|\Psi_E\rangle \langle \Psi_E|\Phi_m\rangle =\delta_{n,m}\,,\quad
   n,m=0,1,\ldots,
 \end{equation}
  where $|\Phi_n\rangle $, $n=0,1,\ldots $ is an orthonormal basis in $H$.
  In coordinate representation the kets $|\Psi_E\rangle $ are reduced
  to the smooth spinors  $\psi_E$. Therefore the action of $L$ on them
  is well-defined. Let us consider the following family of spinors
      \begin{equation} \label{PhiE}
      |\Phi_E\rangle =   N_E^{-1}L|\Psi_E \rangle  \,,\quad \forall
|\Psi_E\rangle \in D,
       \end{equation}
  where $N^2_E=  (E-\lambda_1)(E-\lambda_2)$.
  It is easy to see that these spinors are elements of $H$ if $ N^2_E>0$.
 Indeed, they are eigenspinors of $h_1$: $h_1|\Phi_E\rangle  =E|\Phi_E\rangle
$.
  Therefore, taking into
 account the factorization property (\ref{LPL}),
 we can define the action of $L^+$ on the spinor $|\Phi_E\rangle $
  \begin{equation}
  L^+ |\Phi_E\rangle  =N_E^{-1}L^+ L|\Psi_E\rangle =N_E|\Psi_E\rangle  \,.
   \end{equation}

Remark that the right hand side of (\ref{IP}) is nothing
 but the sum of two integrals with respect to Lebesque measure on
 the real line. Therefore, when $L$ acts on eigenspinors of $h_0$ and $L^+$
 acts on these of $h_1$, the  adjoint operation formally introduced at the begining of
Section 3 coincides
 with the adjoint with respect to the inner product (\ref{IP}).
   As a result, for all $\lambda_{1}<\lambda_{2}$ such that the interval
$(\lambda_{1},\lambda_{2})$ does not contain spectral points of $h_0$, we have
 \begin{equation} \label{LPLP}
 \langle \Phi_E|\Phi_E\rangle =N^{-2}_E  \langle  L\Psi_E|L\Psi_E\rangle  =
 N^{-2}_E \langle  \Psi_E|L^+ L\Psi_E\rangle  =
   \langle  \Psi_E|\Psi_E\rangle  \,.
 \end{equation}
 This result means that if $E>\lambda_{2}$ or $E<\lambda_1<\lambda_{2}$  is such that
$E\in
\mbox{spec}(h_0)$,
 then it also happens that  $E\in \mbox{spec}(h_1)$.
 The inverse statement is in general not valid.
 Hence, to find the whole spectrum of $h_1$ it remains to
 analyse the values $\lambda_1$ and $\lambda_2$ only. Here, several
 possibilities can take place.
Let us consider two real numbers $\tilde\mu$ and $\mu$,  $\tilde\mu<\mu$, such that
the interval $(\tilde\mu,\mu)$ is a spectral gap of $h_0$, for instance the gap between
the positive and negative parts of the spectrum, if any. Then, the main posibilities are
the following:
\begin{itemize}
\item[(I)]
If $\mu$ is a level of the discrete spectrum, then one can take $\lambda_2=\mu$,
$\tilde\mu<\lambda_1<\lambda_2$, and $u_2=\psi_\mu$.
 \begin{itemize}
\item[(A)]
 If, in addition,
 there exists  a linear combination of $v_1$ and
 $\widetilde v_1$,
 $\varphi  =c_1v_1+c_2\widetilde v_1$
 (we recall that $v_1$ ($\widetilde v_1$) is the
 first column of the matrix ${\mathcal V}$ ($\widetilde{\mathcal V}$)),
such that $\varphi  \in H$, then the
 level $E=\mu$ will not be present in the spectrum of $h_1$, but the
 new discrete level $E=\lambda_1$ will appear.
\item[(B)]
 If the last condition is not satisfied,
 the spectrum of $h_1$ will coincide with the spectrum of $h_0$,
 except for one level $E=\mu$, which is missing in the spectrum of $h_1$.
\end{itemize}
 \item[(II)]
If both $\lambda_1$ and $\lambda_2$ fall inside the interval  $(\tilde\mu,\mu)$,
$\tilde\mu<\lambda_1<\lambda_2<\mu$, then the following posibilities arise:
\begin{itemize}
\item[(A)]
The spectrum of $h_1$ totally coincides with the spectrum of $h_0$.
\item[(B)]
 One discrete level is created at $E=\lambda_1$
\item[(C)]
 One discrete level is created at $E=\lambda_2$.
\item[(D)] Two new levels are created at $E=\lambda_1$
 and $E=\lambda_2$.
\end{itemize}
\end{itemize}

 Since the functions $\Psi_E$ are smooth enough, we know the action
 of $L$ not only on these spinors, but also on any  linear
 combination of them of the form
 \begin{equation}\label{finit}
|\Psi\rangle  =\int d\rho (E)C(E)|\Psi_E\rangle  ,
 \end{equation}
where $C(E)$ is a finite function over $\mathbb R$, i.e., a function with a compact
support. The set of spinors (\ref{finit}) is a linear space that we will denote in the
sequel by
${\mathcal L}_0$. We notice that it is dense in $H$. The image of the space ${\mathcal
L}_0$
 under the action of the operator $L$
 consists of the spinors
  \begin{equation}\label{finit1}
 |\Phi\rangle  =\int d\rho (E)N_EC(E)|\Phi_E\rangle  \,.
  \end{equation}
 This new set will be denoted by ${\mathcal L}_1$.
  The operator $L^+$, being a linear operator, is defined for all
  $|\Phi\rangle  \in {\mathcal L}_1$. Moreover, the following equality holds
   \begin{equation}
  \langle  L\Psi|\Phi\rangle  =\langle \Psi |L^+\Phi\rangle \,,\qquad
 \forall |\Psi\rangle \in {\mathcal L}_0\,,\quad \forall |\Phi\rangle \in {\mathcal
L}_1 \,.
   \end{equation}
   Nevertheless, this does not mean that the operator $L^+$ is
   adjoint to $L$ with respect to the inner product in $H$.
   To find such an operator one has to specify correctly its
   domain of definition. We shall not look for this domain.
   Instead we shall give a closed extension $\bar{L}$ of the
   operator $L$ and then find its adjoint $\bar{L}^\dagger$ with respect
   to the inner product in $H$.

 For simplicity, let us suppose that $h_1$ is completely
 isospectral with $h_0$. (If this is not the case, the reasonings are
 similar, but it will be necessary to specify the spaces where the operators act on.)
 In this case the set of functions (\ref{PhiE})
 is another basis in the same space $H$ and the operator
\begin{equation}
U=\int d\rho (E) |\Phi_E\rangle  \langle  \Psi_E |
\end{equation}
 realizes an unitary mapping of $H$ onto itself,
\begin{equation}
 U^{-1}=U^\dagger = \int d\rho (E) |\Psi_E\rangle  \langle  \Phi_E | \,,
 \end{equation}
provided the vectors $|\Psi_E\rangle $ are orthonormal.
Consider now the following operators
\begin{eqnarray}
\bar L  & = & \int d\rho (E) N_E|\Phi_E \rangle \langle \Psi _E|\, ,
\label{Lb} \\ [1ex]
\bar L^\dagger  & = &  \int d\rho (E) N_E|\Psi_E \rangle \langle \Phi _E|\, ,
\label{Lbk}
\end{eqnarray}
It is not difficult to specify their domains of definition. For
this purpose we
introduce the self-adjoint operator $g_0=(h_0-\lambda_1)(h_0-\lambda_2)$,
wich has the following  spectral decomposition
 \begin{equation}
 g_0=\int d\rho(E)N_E^2|\Psi_E\rangle \langle \Psi_E| \,.
 \end{equation}
 We shall suppose the parameters $\lambda_1$ and $\lambda_2$ to be such that
$g_0$ is positive definite. Then, the spectral decomposition of its square root is
 \begin{equation}\label{g012}
 g_0^{1/2}=\int d\rho(E)N_E|\Psi_E\rangle \langle \Psi_E|  \,.
 \end{equation}
  It follows now that
 \begin{equation}
\label{NE2langle}
 \|\bar L\Psi \|^2=\|g_0^{1/2}\Psi \|^2=
 \int d\rho (E) N_E^2|\langle \Psi |\Psi _E\rangle |^2\,.
 \end{equation}
This means that the domain of definition of $\bar L$ coincides
with that of $g_0^{1/2}$ and consists of all $\psi \in H$ such
that the integral in the right-hand side of (\ref{NE2langle})
converges. The domain of definition of $\bar L^\dagger$ coincides
with that of the operator $g_1^{1/2}$ where
 $g_1=(h_1-\lambda_1)(h_1-\lambda_2)$ is also positive definite.

From formulae (\ref{Lb}) and (\ref{Lbk}) it follows
that the operator $\bar L^\dagger$
is the adjoint of $\bar L$ with respect to the inner product,
the domains of definition of $\bar L$ and $\bar L^\dagger$
being well specified. Moreover,
$\bar L^{\dagger \dagger }=\bar L$. This implies \cite{Smirn,RS} that the
operator $\bar L$ is closed. Expressions (\ref{Lb})--(\ref{Lbk})
give quasispectral representations of the closed operators
$\bar L$ and $\bar L^\dagger$.

From  (\ref{Lb})--(\ref{Lbk}) it also follows
that $\bar L\psi _E=L\psi _E=N_E\varphi _E$ and
$\bar L^\dagger\varphi _E=L^+\varphi _E=N_E\psi _E$. This means
that $\bar L$ is the closed extension of the operator $L$, and
$\bar L^\dagger$ is the closed extension of the operator $L^+$,
when the domains ${\mathcal L}_{0}$ and ${\mathcal L}_{1}$ are
taken as their initial domains of definition.

From the spectral decomposition of the operators $g_0^{1/2}$
(\ref{g012}) and $g_1^{1/2}$,
\[
g_1^{1/2}=\int dp N_p|\varphi _p\rangle
\langle \varphi _p| \,,
\]
one obtains the following representations for $\bar L$ and $\bar
L^+$:
\[\bar L=Ug_0^{1/2}=g_1^{1/2}U \, ,\quad
\bar L^\dagger = g_0^{1/2}U^\dagger = U^\dagger g_1^{1/2}\, .
\]
Such representations are known as  {\it polar
decompositions} or
{\it canonical representations} of
closed operators (see for example \cite{RS,DS}).

Let us consider now bounded operators
\[
M=\int d\rho (E) N_E^{-1}|\Phi _p\rangle \langle \Psi _E| \,,\qquad
M^\dagger =\int d\rho (E) N_E^{-1}|\Psi _p\rangle \langle \Phi _E| \,,
\]
defined in $H$. It is not difficult to see that both $M\bar L^\dagger $
and $M^\dagger\bar L $ are unit operators in $H$. Using the
spectral resolutions of the operators $g_0^{-1/2}$ and $g_1^{-1/2}$,
\[
g_0^{-1/2}=
\int d\rho (E)N_E^{-1}|\Psi _E\rangle \langle \Psi _E| \,,
\qquad
g_1^{-1/2}=
\int d\rho (E)N_E^{-1}|\Phi _E\rangle \langle \Phi _E| \,,
\]
 one
derives the polar decompositions of the operators $M$ and $M^\dagger$:
\[
M=Ug_0^{-1/2}=g_1^{-1/2}U\,,
\qquad
M^\dagger = g_0^{-1/2}U^\dagger = U^\dagger g_1^{-1/2}\,.
\]
It is easily seen that these operators factorize the
inverses of $g_0$ and $g_1$: $M^\dagger M=g_0^{-1}$, $MM^\dagger = g_1^{-1}$.

As a final remark of this Section, we would like to notice that
the space $H$ can be obtained as a closure of the
linear space ${\mathcal L}_0$ of all finite linear combinations of the
functions $|\Phi _E\rangle  =L|\Psi _E\rangle $ with respect to the norm
generated by the inner product (\ref{IP}).
 The set of functions of the form
$|\Phi\rangle  =\bar L|\Psi\rangle  $, when $|\Psi \rangle $ runs through the
whole domain of definition of the operator $\bar L$ (i.e., the
domain of definition of the operator $\sqrt
{g_0}$, $D({\sqrt {g_0}})$), can not give the whole space $H$. Nevertheless, if one
defines a new inner product in
 ${\mathcal L}_1$,
$\langle \Phi _a|\Phi _b\rangle _1\equiv \langle L\Psi _a|L\Psi
_b\rangle = \langle \Psi _a|g_0|\Psi _b\rangle $, $|\Psi
_{a,b}\rangle \in {\mathcal L}_0$, $|\Phi _{a,b}\rangle \in {\mathcal L}_1$, then the
closure of ${\mathcal L}_1$ with respect to the norm generated by this
inner product coincides with the set $|\Phi\rangle  =\bar L|\Psi\rangle  $,
$|\Psi \rangle \in D({\sqrt {g_0}})$. This space is embedded in $H$.

\section{Supersymmetry}

As it has been mentioned in the Introduction,
supersymmetric transformations in the
non-relativistic quantum mechanics are basically equivalent to
Darboux transformations for the Schr\"odinger equation. This
equivalence is based on two properties of the Darboux
transformation: (a) the intertwining relations
and (b) the factorization of the Hamiltonians,
similar to that established in Section~4.

In the case of the Dirac equation we started from the intertwining
relation and proved the factorization properties.
Therefore, every property giving rise to the supersymmetry
of the Schr\"odinger equation also takes place for the Dirac equation. Moreover, the
transformation operators are well-defined in the Hilbert space
$L^2({\mathbb R})\otimes {\mathbb  C}^2$. Therefore, in order
to show the supersymmetric character of our approach we can proceed in the
same way as in the nonrelativistic case.

To begin, let us introduce the following $4\times 4$ matrices
\begin{equation}
 {\mathcal H} \equiv \left( \begin{array}{cc}
h_0 & 0\\
0 & h_1\\
\end{array}\right)\, ,\quad
{\mathcal Q }^{+}=\left( \begin{array}{cc}
  0 & L^{+}\\
0 & 0\\
\end{array}\right)\, ,\quad
 {\mathcal Q }=\left( \begin{array}{cc}
  0 & 0\\
L & 0\\
\end{array}\right) \, .
\label{eq33}
\end{equation}
It is easily seen that, in one side, the two commutation relations
\begin{equation} \label{eq34}
 [{\mathcal Q },{\mathcal H}]=[{\mathcal Q }^{+},{\mathcal H}]=0
\end{equation}
 are equivalent to the intertwining relations  (\ref{splet}) and
 (\ref{spletsopr}).
  On the other side, the anticommutation relations
  \begin{equation}
  \{ {\mathcal Q } ,{\mathcal Q }^{+} \} \equiv
  {\mathcal Q }{\mathcal Q }^{+}+{\mathcal Q }^{+}{\mathcal Q }=({\mathcal H} -\lambda_1 I )({\mathcal
H} -\lambda_2 I )
  \end{equation}
  are equivalent to the factorizations  (\ref{LPL}) and
  (\ref{LLp}), while
  \begin{equation}
  {\mathcal Q }^2=({\mathcal Q }^{+})^2=0  \,.
  \label{eq35}
  \end{equation}

Remark that relations~(\ref{eq34})--(\ref{eq35}) are those of a quadratic
deformation of the superalgebra $sqm(2)$, inherent to the usual supersymmetric
 quantum mechanics \cite{Wit}. This quadratic superalgebra cannot be seen
 directly from the Dirac equation, and therefore we associate it
 with a hidden supersymmetry. Let us also point out that
 a superalgebra similar to that of (\ref{eq34}-\ref{eq35}) can also be found
 in the non-relativistic context, when second order
 Darboux transformations are considered \cite{10,10a}.

 \section{Chains of Darboux transformations}

The aim of this Section is to iterate the one-step transformations we have considered
in previous sections. In order to accomplish this goal, we will rewrite first the
formulas (\ref{eql}) and (\ref{V1}) in a form wich will be more appropriate for this
purpose. Remark that the action of the operator
$L$ to $\psi $ can be rewritten as follows
\begin{equation} \label{phi_diff}
\varphi = L \psi = \psi_x - {\mathcal U}_x\,  {\mathcal U}^{-1} \psi =
 {\mathcal U} ({\mathcal U}^{-1} \psi)_x \, .
\end{equation}
Now, we will substitute the term in parentheses by the following expression
\begin{equation}
{\mathcal U}^{-1} \psi = \frac{1}{W(f,g)} \left(
\begin{array}{l} W(\psi,g) \\ [1ex]
W(f, \psi) \end{array} \right),
\label{u-1det}
\end{equation}
where  $W(f,g) = f^t \gamma g = f_1 g_2 - f_2 g_1$ is the
Wronskian for the Dirac equation that we introduced in Section~4, being $f$ and $g$
the spinors from wich the matrix ${\mathcal U}$ is composed: ${\mathcal U}=(f,g)$,
$f=(f_1,f_2)^t$, $g=(g_1,g_2)^t$, and $h_0f=\lambda f, h_0 g=\mu g$, i.e.,
$\Lambda={\rm diag}(\lambda,\mu)$. It can be established by a straightforward
calculation that the derivative of (\ref{u-1det}) can be written as
\begin{equation}
({\mathcal U}^{-1} \psi)_x = \frac{1}{[W(f,g)]^2} \left(
\begin{array}{l} W_S(W(f,g), W(\psi,g))   \\ [1ex]
W_S(W(f,g), W(f, \psi)) \end{array} \right),
\label{u-1xdet}
\end{equation}
were $W_S(f,g) = f g' - f' g$ is the usual Wronskian.
In the sequel we will use both, the prime
and the subscript $x$, for denoting the derivative with respect to $x$; as usual, a
multiple derivative of a function $f$ will be denoted as $f^{(n)}$.

After replacing
$({\mathcal U}^{-1} \psi)_x$ for its expression in (\ref{u-1xdet}), one gets from 
(\ref{phi_diff}) the desired  form for
$\varphi $:
\begin{equation}\label{phi_det}
\varphi = \frac{1}{W(f,g)}
 \left(   \hspace{0.1em}
\left|
\begin{array}{lll} f_1 & g_1 & \psi_1  \\ f_2 & g_2 & \psi_2 \\
 f'_1 & g'_1 & \psi'_1
\end{array}
\right|, \hspace{0.3em}
\left|
\begin{array}{lll} f_1 & g_1 & \psi_1  \\ f_2 & g_2 & \psi_2 \\
 f'_2 & g'_2 & \psi'_2
\end{array} \right|  \hspace{0.1em}
\right)^t .
\end{equation}
The symbols $\left|\begin{array}{cc} \cdot & \cdot \\ \cdot &
\cdot \end{array}\right| $
are used  to denote determinants.
A similar calculation results in another representation of
 the transformed potential matrix $V_1$ given in (\ref{V1}):
\begin{equation}
V_1 = V_0 + [\gamma , D_1],
\label{v-1det}
\end{equation}
where the square brackets represent the commutator. The matrix $D_1$ is
\begin{equation}
D_1 = \frac{1}{W(f,g)} \left(
\begin{array}{ll}
\left| \begin{array}{ll} f'_1 & g'_1 \\ f_2 & g_2 \end{array}
\right| &
\left| \begin{array}{ll} f_1 & g_1 \\ f'_1 & g'_1 \end{array} \right| \\
[3ex]
\left| \begin{array}{ll} f'_2 & g'_2 \\ f_2 & g_2 \end{array}
\right| & \left| \begin{array}{ll} f_1 & g_1 \\ f'_2 & g'_2
\end{array} \right|
\end{array} \right).
\label{D-det}
\end{equation}
In the next subsections we generalize the formulae (\ref{phi_det})--(\ref{D-det}) to a
chain of $n$ consecutive transformations of the same type.

\subsection{Transformation of spinors}

Let us denote now by $L_{1\leftarrow 0}$ the first order operator intertwining $h_0$
and $h_1$, as given in  (\ref{eql}), having the transformation
function ${\mathcal U}_1$ with the eigenvalue $\Lambda_1 = \mbox{diag}
(\lambda_1, \mu_1)$, $\lambda_1\ne \mu_1$.
Let ${\mathcal U}_2$ be a solution of (\ref{dirac0}) with
$\Lambda_2 = \mbox{diag} (\lambda_2,\mu_2)$,
 $\lambda_2\ne \mu_2$,
$\left\{\lambda_2, \mu_2\right\}\ne \left\{\lambda_1,\mu_1\right\}$.
The last condition, together with the
non-degeneracy of ${\mathcal U}_1$, means the non-degeneracy of the matrix
${\mathcal V}_2 = L_{1\leftarrow 0}{\mathcal U}_2$.
The function ${\mathcal V}_2$ being an eigenfunction of $h_1$ may be
chosen as the transformation function for the second transformation step.

The operator carrying out
the second transformation will be denoted by $L_{2\leftarrow 1}$. It
intertwines $h_1$ and $h_2$, the new potential given by the
formula (\ref{V1}) in which the replacements $V_{0}\to V_1$, $V_{1}\to V_2$ and
${\mathcal U}\to{\mathcal V}_2$ are done. It follows from here that the second order
operator
\begin{equation}
L_{2\leftarrow 0} = L_{2\leftarrow 1} L_{1\leftarrow 0} \label{L-02}
\end{equation}
intertwines $h_0$ and $h_2$ and realizes the transformation from
$h_0$ directly to $h_2$, without using an intermediate potential
$V_1$. It is completely defined by two transformation functions
${\mathcal U}_1$ and ${\mathcal U}_2$. It is clear that subsequent iterations give
us an $n$th order transformation operator
\begin{equation}  \label{L-0n}
L_{n\leftarrow 0} = L_{n\leftarrow n-1} \cdots L_{2\leftarrow 1} L_{1\leftarrow 0}\,,
\end{equation}
realizing the transformation between the first $h_0$ and the last
$h_n$ elements of the chain of Hamiltonians $h_0, h_1,\ldots ,h_n$.
The operator $L_{n\leftarrow 0}$ is defined by $n$ transformation functions
${\mathcal U}_1,\ldots ,{\mathcal U}_n$
\begin{equation}
\label{dirac-ui}
h_0 {\mathcal U}_i = {\mathcal U}_i \Lambda_i, \quad \Lambda_i =
\mbox{diag} (\lambda_i, \mu_i)\,.
\end{equation}
We shall derive now a compact formula for the eigenfunctions of the
Hamiltonian $h_n$ (not necessary belonging to the Hilbert space). For
the columns of the matrix ${\mathcal U}_i$ we introduce the notations $f_i =
(f_{i1}, f_{i2})^t$, $g_i = (g_{i1}, g_{i2})^t$, so that
${\mathcal U}_i=(f_i\,, g_i)$, $f_i$, $g_i$ being two-component
vector-columns (spinors), and $h_0f_i=\lambda_i f_i$, $h_0g_i=\mu_i g_i$.

In order to obtain the formulas we are looking for, we need to introduce new
notations. Let $W(f_1,$ $g_1, \ldots, f_n,g_n)$ be a $(2n)$th order
determinant organized as follows. The first two rows of this
determinant are composed of the components of the spinors  $f_1$,
$g_1$, \ldots, $f_n$, $g_n$, the first line from the first
components and the second line from the second components; any
subsequent pairs of rows is the derivative of the previous pair.
 Thus, we can write
\begin{equation}  \label{w-def}
W(f_1,g_1,\ldots, f_n,g_n) \equiv W(f_1,\ldots, g_n)= \det(a_{i,j}), \  i,j =
1,2, \ldots 2n \,,
 \end{equation}
where
 \begin{equation}
\left. \begin{array}{ll}
a_{2l-1,2m-1} = f^{(l-1)}_{m1}, & a_{2l,2m-1} =
f^{(l-1)}_{m2}, \\ [2ex]
a_{2l-1,2m} = g^{(l-1)}_{m1}, & a_{2l,2m} = g^{(l-1)}_{m2},
 \end{array} \right\}
\label{aijaij}
\end{equation}
$l,m = 1, \ldots, n$. Remember  that $W(f_1,g_1,\ldots, f_n,g_n)$  is the determinant of
an even order matrix.

 We need also some odd order determinants. If we have the same set of $2n$ spinors
$f_i$ and $g_i$, $i=1,\ldots ,n$, plus an additional spinor  $\psi =(\psi_1,\psi_2)^t$,
then we can construct two different determinants of $(2n+1)$th order
from the previous determinant (\ref{w-def}): by adding a column composed
 of these spinors (using the  procedure described above) and a row
 composed of the $n$th derivative of either the upper elements of the
 spinors $f_1$, $g_1$, \ldots, $f_n$, $g_n$, $\psi$, or the lower
 elements of the spinors. In this way, we can get determinants of two kinds
 \begin{eqnarray}
\label{wronski1}
 W_1(f_1,g_1,\ldots, f_n,g_n,\psi)  &  = &  \det(b_{i,j})\,,
 \label{w1-def} \\ [1ex]
\label{wronski2}
 W_2(f_1,g_1,\ldots, f_n,g_n,\psi)  &  = &  \det(c_{i,j})\,, \quad
 i,j = 1,2, \ldots 2n+1 \,,
\end{eqnarray}
where $b_{i,j} = c_{i,j} = a_{i,j}$ when $i,j = 1,\ldots,2n$, and
\begin{equation}
\left. \begin{array}{l}
b_{2l-1,2n+1} = c_{2l-1,2n+1} = \psi^{(l-1)}_{1},  \ l = 1, \ldots, n; \\ [1ex]
b_{2l,2n+1} = c_{2l,2n+1} = \psi^{(l-1)}_{2}, \ l = 1, \ldots, n; \\ [1ex]
b_{2n+1,2m-1} = f^{(n)}_{m1}, \
c_{2n+1,2m-1} = f^{(n)}_{m2},  \ m =1, \ldots, n; \\ [1ex]
b_{2n+1,2m} = g^{(n)}_{m1}, \
c_{2n+1,2m-1} = g^{(n)}_{m2},  \ m =1, \ldots, n; \\ [1ex]
b_{2n+1,2n+1} = \psi^{(n)}_{1}, \  c_{2n+1,2n+1} = \psi^{(n)}_{2}.
 \end{array} \right\}
\label{wronski3}
\end{equation}

Now, in order to prove the main result of this Section, we need the
following auxiliary statements.
\begin{lemma}
\label{Lm3}
The following equation takes place:
\begin{eqnarray}
\hskip-0.5cm
&&
W_S \bigl( W(f_1,g_1, \ldots , f_n,g_n), W(f_1,g_1, \ldots, f_n,\psi) \bigr)
=\nonumber
\\ [1ex]
\hskip-0.5cm
 &&
\hskip2cm  = W_1(f_1,g_1, \ldots , f_{n-1},g_{n-1},f_n) W_2(f_1,g_1, \ldots ,
f_n,g_n,\psi) \label{diff-w2n}
\\ [1ex]
\hskip-0.5cm
&&
\hskip2.1cm  - W_2(f_1,g_1, \ldots , f_{n-1},g_{n-1},f_n) W_1(f_1,g_1,
\ldots, f_n,g_n,\psi) \, . \nonumber
\end{eqnarray}
\end{lemma}

{\it Proof.}
To prove the lemma we need first the following identity,
which can be checked in \cite{Sall-87}:
\begin{equation}
W^\ell_{\ell \ell} W^\ell_{jk} - W^\ell_{j \ell}
W^\ell_{\ell k} = W^{\ell+1}_{jk} W_{\ell -1},
\label{hartman}
\end{equation}
where
\begin{eqnarray*}
&&
W^\ell_{jk}:=W_{1,2,\ldots,\ell-1,j;1,2,\ldots,\ell-1,k},\quad j,k >\ell,\\ [1ex]
&&
W_{i_1,i_2,\ldots,i_k,;j_1,j_2,\ldots,j_k} := \det(y_{i_m,j_n}), \quad m,n = 1,\ldots,k,
\end{eqnarray*}
and $y_{i_m,j_n}$ are arbitrary.

Note first that the derivative of a determinant such as the one
defined in
 (\ref{w-def}) is the sum of two determinants of a similar structure.
 The difference is only either in the last row or in the next-to-last
 row.
 In the first case
 we obtain the determinant denoted by   $R_1(f_1,g_1,\ldots, f_n,g_n)$
 in which
  the last row is replaced by the $n$th derivative
 of the second elements of the spinors
 $f_1$, $g_1$, \ldots, $f_n$, $g_n$
 and in the second case we get the determinant denoted by
  $R_2(f_1,g_1,\ldots, f_n,g_n)$ in which
 the
 next-to-last row is replaced by the $n$th derivative of the first
 elements of the same spinors.
To be more precise, we have the following
\begin{equation}
 [W(f_1, \ldots,  g_n)]' =
R_1(f_1,g_1,\ldots, f_n,g_n) +
R_2(f_1,g_1,\ldots, f_n,g_n) \,, \label{w-diff}
\end{equation}
where
\begin{eqnarray}
&&
R_1(f_1,g_1,\ldots, f_n,g_n) = \det(r^{1}_{i,j})\,,
\label{R1-def} \\ [1ex]
&&
R_2(f_1,g_1,\ldots, f_n,g_n) = \det(r^{2}_{i,j})\,,
\quad i,j = 1,2, \ldots 2n \,, \label{R2-def}
\end{eqnarray}
and
 \begin{equation}
\left. \begin{array}{l}
r^{1}_{i,j} = r^{2}_{i,j} = a_{i,j}, \ \ i = 1, \ldots,
2n-2,\  j = 1, \ldots, 2n;
\\ [1ex]
r^{1}_{2n-1,2m-1} = f^{(n)}_{m1},\ r^{1}_{2n-1,2m} = g^{(n)}_{m1}, \ \ m = 1, \ldots,
n;
\\ [1ex]
r^{2}_{2n,2m-1} = f^{(n)}_{m2},\  r^{2}_{2n,2m} = g^{(n)}_{m2}, \ \ m = 1, \ldots,
n;
\\ [1ex]
r^{1}_{2n,j} = a_{2n,j},\  r^{2}_{2n-1,j} = a_{2n-1,j}, \ \ j = 1, \ldots, 2n.
 \end{array} \right\}
\label{2aijaij}
\end{equation}
The functions $a_{i,j}$ are defined in (\ref{aijaij}). The Lemma  follows now from
(\ref{hartman}) and (\ref{w-diff}).
\hfill$\Box$

\begin{proposition}
In what follows, we will need also the following identity
\begin{eqnarray}
&&
W_1(f_1,g_1,\ldots, f_{n-1},g_{n-1},f_n) W_2(f_1,g_1,\ldots, f_{n-1},g_{n-1},g_n)-
\nonumber  \\ [1ex]
&&
\label{w2n+1-jak}
\qquad
-W_1(f_1,g_1,\ldots, f_{n-1},g_{n-1},g_n)W_2(f_1,g_1,\ldots, f_{n-1},g_{n-1},f_n) \\
[1ex] &&
 \quad = W(f_1,g_1, \ldots, f_{n-1},g_{n-1})\ W(f_1,g_1, \ldots, f_n,g_n) \,.
\nonumber
\end{eqnarray}
which is a direct consequence of {\rm (\ref{hartman})}.
\end{proposition}

Now we formulate and prove the main result of this Section.
\begin{theorem}
The action of the operator $L_{n\leftarrow 0}$ on a function
 $\psi = (\psi_1,\psi_2)^t$ reads as follows
\begin{equation}
\label{L0n-psi}
L_{n\leftarrow 0} \psi = \frac{1}{W(f_1,g_1,\ldots,f_n,g_n)} \left(
\begin{array}{l}
W_1(f_1,g_1,\ldots,f_n,g_n,\psi) \\ [1ex]
W_2(f_1,g_1,\ldots,f_n,g_n,\psi) \end{array} \right)\,,
 \end{equation}
where $W_1$ and $W_2$ are defined by {\rm (\ref{wronski1})--(\ref{wronski3})}.
\end{theorem}

{\it Proof.}
To prove the theorem we use the  perfect
induction method. Let us suppose that the action of the operator
$L_{n-1\leftarrow 0}$ on a function $\psi $
 have the form (\ref{L0n-psi}), with the replacement $n\to n-1$, i.e.,
\begin{equation}
\label{psi-tilde}
\widetilde{\psi} = L_{n-1\leftarrow 0} \psi =
\frac{1}{W(f_1,\ldots ,g_{n-1})} \left(
\begin{array}{l}
W_1(f_1,g_1,\ldots,f_{n-1},g_{n-1},\psi) \\ [1ex]
W_2(f_1,g_1,\ldots,f_{n-1},g_{n-1},\psi) \end{array} \right)\,.
\end{equation}
Then, according to  (\ref{L-0n})  and (\ref{phi_diff}), we have
\begin{equation}
L_{n\leftarrow 0} \psi = L_{n-1\leftarrow n} \widetilde{\psi} =
\widetilde{{\mathcal U}}_n (\widetilde{{\mathcal U}}_n^{-1}\widetilde{\psi} )_x ,
\label{psi-ut}
\end{equation}
where
\begin{equation}  \label{u-tilde}
\widetilde{{\mathcal U}}_n = L_{n-1\leftarrow 0} {\mathcal U}_n\,,  \quad
\widetilde{{\mathcal U}}_n = (\widetilde{f}_n,\widetilde{g}_n) \,.
\end{equation}
Here,  the spinors $\widetilde{f}_n$ and $\widetilde{g}_n$ should be
calculated by the same formula  (\ref{psi-tilde}), where $\psi $ has to be
replaced by $f_n$ and $g_n$, respectively. Using Eq.~(\ref{u-1det}) we find
 \begin{equation}
\label{ut-1det}
 \widetilde{{\mathcal U}}_n^{-1}\widetilde{\psi} =
 \frac{1}{W(\widetilde{f}_n,\widetilde{g}_n)} \left(
 \begin{array}{l}
 W(\widetilde{\psi},\widetilde{g}_n) \\  [1ex]
 W(\widetilde{f}_n,\widetilde{\psi}) \end{array} \right)\,.
 \end{equation}
After calculating those Wronskians, and using (\ref{psi-tilde}) and
(\ref{w2n+1-jak}), we get from (\ref{ut-1det}) the equation
\begin{equation} \label{ut-1det1}
\widetilde{{\mathcal U}}_n^{-1}\widetilde{\psi} =
\frac{1}{W(f_1,g_1,\ldots,f_n,g_n)} \left(
\begin{array}{r}
- W(f_1,g_1,\ldots,g_n,\psi)\\ [1ex]
 W(f_1,g_1,\ldots,f_n,\psi)\end{array} \right).
 \end{equation}
The derivative of this function reads
$$
 (\widetilde{{\mathcal U}}_n^{-1}\widetilde{\psi} )_x =
\frac{1}{[W(f_1,\ldots,g_n)]^2} \left(
\begin{array}{r}
- W_S(W(f_1,\ldots,g_n),W(f_1,\ldots,g_n,\psi)) \\ [1ex]
 W_S(W(f_1,\ldots ,g_n),W(f_1,\ldots,f_n,\psi))
\end{array} \right).
$$
The statement of the Theorem is a direct implication of last equation, together with
equations (\ref{diff-w2n}), and (\ref{psi-ut}).
\hfill$\Box$

\subsection{Transformation of the potential}

In the last part of this Section we need to introduce the following notation: let us
denote by $Q_1(f_1,g_1,\ldots,f_n,g_n)$ the determinant constructed from
$W(f_1,g_1,\ldots,f_n,g_n)$ by the replacement of the $(2n)$th line
with the $n$th derivatives of the first elements of the spinors
  $f_1$, $g_1$, \ldots, $f_n$, $g_n$, and let us denote by $Q_2(f_1,g_1,\ldots,f_n,g_n)$
 the determinant obtained from
 $W(f_1,g_1,\ldots,f_n,g_n)$
 by the replacement of $(2n-1)$th line with the $n$th derivatives of
 of the second elements of the same spinors. Thus, we have
 \begin{eqnarray}
  Q_1(f_1,g_1,\ldots, f_n,g_n)   &   =  &    \det(q^{1}_{i,j})\,,
 \label{Q1-def} \\ [1ex]
  Q_2(f_1,g_1,\ldots, f_n,g_n)   &   =  &    \det(q^{2}_{i,j})\,, \quad
 i,j = 1,2, \ldots 2n \,,
 \label{Q2-def}
 \end{eqnarray}
where
 \begin{equation}
\left. \begin{array}{l}
 q^{1}_{i,j} = q^{2}_{i,j} = a_{i,j}, \ i = 1, \ldots,
2n-2,\ j = 1, \ldots, 2n;  \\  [1ex]
 q^{1}_{2n,2m-1} = f^{(n)}_{m1}, \ q^{1}_{2n,2m} = g^{(n)}_{m1}, \
m = 1, \ldots, n;  \\   [1ex]
q^{2}_{2n-1,2m-1} = f^{(n)}_{m2},\ q^{2}_{2n-1,2m} = g^{(n)}_{m2},\
m = 1, \ldots, n;  \\   [1ex]
 q^{1}_{2n-1,j} = a_{2n,j}, \ q^{2}_{2n,j} = a_{2n-1,j}, \ j = 1, \ldots, 2n,
 \end{array} \right\}
\label{qijqij}
\end{equation}
and the functions $a_{i,j}$ are defined in (\ref{aijaij}).

\begin{theorem}
The potential resulting from a chain of Darboux transformations
{\rm (\ref{L-0n})}
 has the form
 \begin{equation}  \label{vn-det}
 V_n = V_0 + [\gamma , D_n]\,,
  \end{equation}
where
 \begin{equation}    \label{Dn-det}
 D_n = \frac{1}{W(f_1, \ldots, g_n)} \left(
 \begin{array}{ll}
 R_1(f_1,g_1,\ldots,f_n,g_n) & Q_1(f_1,g_1,\ldots,f_n,g_n) \\ [1ex]
 Q_2(f_1,g_1,\ldots,f_n,g_n) & R_2(f_1,g_1,\ldots,f_n,g_n)
 \end{array} \right),
  \end{equation}
being $R_1$ and $R_2$ defined by {\rm (\ref{R1-def})--(\ref{2aijaij})}, and
$Q_1$ and $Q_2$ defined by {\rm (\ref{Q1-def})--(\ref{qijqij})}.
 \end{theorem}

{\it Proof.}
 To prove this Theorem  we use again the perfect induction method.
 Therefore, let us suppose that the formula (\ref{vn-det}) is valid for the potential
 $V_{n-1}$. Then,  Eq.~(\ref{V1}) implies
\begin{equation}  \label{vn-vn-1}
V_n = V_{n-1} + [\gamma , (\widetilde{{\mathcal U}}_n^{-1})_x\,
\widetilde{{\mathcal U}}_n^{-1}] = V_0 + [\gamma , D_{n-1} +
(\widetilde{{\mathcal U}}_n^{-1})_x\,  \widetilde{{\mathcal U}}_n^{-1} ]\,.
\end{equation}
Here the matrix $\widetilde{{\mathcal U}}_n$ defined in (\ref{u-tilde}) has
the form
\begin{eqnarray}
\widetilde{{\mathcal U}}_n   &   =  &  
\widetilde{w}_nW^{-1}(f_1,g_1,\ldots,f_{n-1},g_{n-1}) ,
 \\  [2ex]
\widetilde{w}_n  &  = &   \left(
\begin{array}{ll}
W_1(f_1,g_1,\ldots,f_{n-1},g_{n-1},f_n) &
W_1(f_1,g_1,\ldots,f_{n-1},g_{n-1},g_n) \\  [1ex]
W_2(f_1,g_1,\ldots,f_{n-1},g_{n-1},f_n) &
W_2(f_1,g_1,\ldots,f_{n-1},g_{n-1},g_n)
\end{array} \right).
\end{eqnarray}
The derivative of this function reads
\begin{eqnarray}
(\widetilde{{\mathcal U}}_n )_x &=&
-\frac{[W(f_1,g_1,\ldots,f_{n-1},g_{n-1})]'}{W(f_1,g_1,\ldots,f_{n-1},g_{n-1})}\
\widetilde{{\mathcal U}}_n  \nonumber \\
&& + \frac{1}{W(f_1,g_1,\ldots,f_{n-1},g_{n-1})}
(\widetilde{w}_n)_x,
\end{eqnarray}
and hence
\begin{eqnarray}  \label{ut_xut-1}
(\widetilde{{\mathcal U}}_n )_x\,  \widetilde{{\mathcal U}}_n ^{-1} &=&
-\frac{[W(f_1,g_1,\ldots,f_{n-1},g_{n-1})]'}{W(f_1,g_1,\ldots,f_{n-1},g_{n-1})}\ I \\
&& + \frac{1}{W(f_1,g_1,\ldots,f_{n-1},g_{n-1})}
(\widetilde{w}_n)_x\,  \widetilde{{\mathcal U}}_n ^{-1}. \nonumber
\end{eqnarray}
After calculating the derivatives and using the same technique as
while proving the Lemma~\ref{Lm3} we obtain
\begin{eqnarray}
 &&\hskip-0.7cm D_n = D_{n-1} + (\widetilde{{\mathcal U}}_n^{-1})_x\,
\widetilde{{\mathcal U}}_n^{-1} =
\nonumber
\\ [1ex]
 &  = &  \frac{1}{W(f_1,\ldots,g_{n-1})} \left(
\begin{array}{ll}
R_1(f_1,g_1,\ldots,f_{n-1},g_{n-1}) & Q_1(f_1,g_1,\ldots,f_{n-1},g_{n-1}) \\  [1ex]
Q_2(f_1,g_1,\ldots,f_{n-1},g_{n-1}) &
R_2(f_1,g_1,\ldots,f_{n-1},g_{n-1})
\end{array} \right) \nonumber
\\  [1ex]
 &- & 
\frac{[W(f_1,\ldots, g_{n-1})]'}{W(f_1,\ldots,g_{n-1})}\ I
\nonumber
\\  [1ex]
 & +&   \frac{1}{W(f_1, \ldots, g_{n-1})} \left( \hspace{-.5em}
\begin{array}{ll}
R_2(f_1,g_1,\ldots,f_{n-1},g_{n-1}) &\hspace{-.5em}
-Q_1(f_1,g_1,\ldots,f_{n-1},g_{n-1}) \\  [1ex]
- Q_2(f_1,g_1,\ldots,f_{n-1},g_{n-1}) &\hspace{-.5em}
R_1(f_1,g_1,\ldots,f_{n-1},g_{n-1})
\end{array} \hspace{-.6em}\right)
\nonumber
\\  [1ex]
 & +&  \frac{1}{W(f_1,\ldots,g_n)} \left(
\begin{array}{ll}
R_1(f_1,g_1,\ldots,f_n,g_n) & Q_1(f_1,g_1,\ldots,f_n,g_n) \\  [1ex]
Q_2(f_1,g_1,\ldots,f_n,g_n) & R_2(f_1,g_1,\ldots,f_n,g_n)
\end{array} \right).
\label{BigFormula}
\end{eqnarray}
The sum of the first and the third terms in this expression
gives
$$
[R_1(f_1,g_1,\ldots,f_{n-1},g_{n-1}) +
R_2(f_1,g_1,\ldots,f_{n-1},g_{n-1})]/
W(f_1, \ldots, g_{n-1})\ I.
$$
According to (\ref{w-diff}),
this is equal to the derivative of
$W(f_1, \ldots, g_{n-1})$, and hence
these items cancel out when
added to the second term of (\ref{BigFormula}). Therefore, we get exactly the
formula (\ref{Dn-det}).
\hfill$\Box$

 As a final remark of this Section, we notice that the formulae
 (\ref{L0n-psi}) and (\ref{vn-det})--(\ref{Dn-det})
 can be considered as relativistic analogs of the
  Crum determinant formulae \cite{Crum},
a result which is very well-known in the non-relativistic case.
Moreover, since they present the final result of
the action of a chain of first order transformations, the operator $L_{n\leftarrow 0}$
and its formally adjoint  $L_{n\leftarrow 0}^+$ satisfy the following
 factorization properties:
 \begin{eqnarray}
 L_{n\leftarrow 0}^+L_{n\leftarrow 0} &=& \prod_{j=1}^n(h_0-\lambda_j)(h_0-\mu_j)\,,
 \\
L_{n\leftarrow 0}L_{n\leftarrow 0}^+ &=& \prod_{j=1}^n(h_n-\lambda_j)(h_n-\mu_j)\,.
\end{eqnarray}
This result comes out directly from Theorem~\ref{th1}.

\section{Pseudoscalar potentials}

\subsection{Darboux transformation for a pseudoscalar potential}

A general pseudoscalar potential is defined only by one function $q_0(x)$,
$x\in \mathbb R$:
\begin{equation}\label{pseudo}
V_0 = m \sigma_3 + q_0(x) \, \sigma_1 =
\left(\begin{array}{cc} m & q_0(x)
\\ [1ex]
q_0(x) & -m
\end{array}\right) ,
\end{equation}
where $m$ is the mass of the particle. The Dirac system for the
components of the spinor $\psi =(\psi_1,\psi_2)^t$ is
\begin{eqnarray} \label{dirac-c}
- \psi'_1 + q_0 \psi_1  &  = &  (E + m) \psi_2\,, \\  [1ex]
 \psi'_2 + q_0 \psi_2  &  = &  (E - m) \psi_1 \ .
\end{eqnarray}
 We would like to notice the following property of this system, that we will
 use in the sequel: if one of the components of the given spinor
 $\psi $ is zero, then the other is also zero for all values of $E$, except if
 $E=\pm m$. When $E=m$ the system has a solution of the form
 $\psi = (\psi_1,\, 0)^t$, and when $E=-m$ the solution is $\psi = (0, \, \psi_2)^t$.

  In general, after applying the Darboux transformation
to a pseudoscalar    potential we get a potential which is not
pseudoscalar anymore. Here we shall formulate
 additional conditions for Darboux transformations to keep
  pseudoscalarity of a potential.

 It is easy to see that if one of the elements of a
 transformation function is zero, then the value $d_1$ defined by
 (\ref{d12}) is constant, $d_1=\pm (\lambda_1-\lambda_2)$. This
 means that after such a transformation the new potential is also pseudoscalar, the
role of the mass being played by $d_1-m$.

 Let us take one of the component of the spinor $u_1$
equal to zero, for instance $u_{21}=0$. This is possible for
 $\lambda_1=m$. In this case
 \begin{equation}  \label{uPS}
  {\mathcal U} =
  \left(\begin{array}{lr}
  u_{11} & u_{12}\\
  0       & u_{22}
  \end{array}\right)\,,
 \end{equation}
 $\det {\mathcal U} = u_{11}u_{22}$, and the potential $V_1$, given in
(\ref{V1D}), takes the
 form
\begin{equation}\label{VV11}
 V_1 =-\lambda_2 \sigma_3 +
 \left[(\lambda_2 - m) \frac{u_{12}}{u_{22}} - q_0\right]\sigma_1\,.
\end{equation}
Now, from the Dirac system we can find  the function $u_{12}$
\begin{equation}
 u_{12} = \frac{1}{\lambda_2 - m}(u'_{22} + q_0 u_{22})\, ,
 \label{u12f}
 \end{equation}
and rewrite Eq.~(\ref{VV11}) as follows
 \begin{equation} \label{ps_V1}
 V_1  = - \lambda_2 \sigma_3 + q_1 \sigma_1\, ,
 \end{equation}
where
\begin{equation} \label{ps_v1}
q_1  = \frac{u'_{22}}{u_{22}} = (\ln u_{22})'\,.
\end{equation}
 In the transformed Dirac system the role of the mass is played
 by $-\lambda_2$.
 We would like also to mention
a relationship existing between the nonzero component of the spinor $u_1$ and the
potential $q_0$,  which immediately follows from the initial Dirac system
 (\ref{dirac-c}):
 \begin{equation}   \label{ps_v0}
 q_0 = \frac{u'_{11}}{u_{11}} \,.
 \end{equation}

Let us find now solutions of the transformed equation. We first
calculate the product ${\mathcal U}_x\, {\mathcal U}^{-1}$:
$$
{\mathcal U}_x\,  {\mathcal U}^{-1} =\left(\begin{array}{cc} (\ln u_{11})' &
\displaystyle{\frac{-u'_{11}u_{12} + u'_{12}u_{11}}{u_{11}u_{22}}
\vphantom{\frac{\frac{x_2}{x_1}}{ \frac{x_2}{x_1}}}} \\  [2ex]
0 & (\ln u_{22})'
\end{array}\right).
$$
Then, we simplify this expression with the help of equations
(\ref{u12f}), (\ref{ps_v1}) and (\ref{ps_v0}):
$$
{\mathcal U}_x\,  {\mathcal U}^{-1} =\left(\begin{array}{cc}
(\ln u_{11})' & -(\lambda_2 + m) \\  [1ex]
0 & (\ln u_{22})'
\end{array}\right) =
\left(\begin{array}{cc}
q_0 & -(\lambda_2 + m) \\  [1ex]
0 & q_1
\end{array}\right)\,.
$$
Finally, using (\ref{eql}), we find the action of the operator $L$ on solutions of the
 initial equation:
 \begin{equation}
 \varphi = L \psi =  \left(\begin{array}{c}
  (\lambda_2 - E) \psi_2\\  [1ex]
  \psi'_2 - (\ln u_{22})' \psi_2
 \end{array}\right).
 \label{ps_phi}
 \end{equation}
We observe that the lower component of this spinor is
 defined just by the same expression that appears in the Darboux
 transformation for the Schr\"odinger equation  with the
 transformation function $u_{22}$
 (see e.g. \cite{BSTMF}).
 We also notice that the upper  component
 of the spinor (\ref{ps_phi})
 differs from $\psi_2$ only by a constant factor.
 This means that they should satisfy the same equation. Later, we
 shall show that this is really the case.

 For the case $u_{11} = 0$ and $\lambda = -m$, similar calculations give us the
following result:
  \begin{eqnarray}
  V_1  &  = &   \lambda_2 \sigma_3 + q_1 \sigma_1\,,
  \label{ps_V1a} \\  [1ex]
  q_1  &  = &  -(\ln u_{12})',
  \label{ps_v1a} \\  [1ex]
  \varphi  &  = & 
\left(\begin{array}{c}
   \psi'_1 - (\ln u_{12})' \psi_1  \\  [1ex]
   (\lambda_2 + E)\psi_1
  \end{array}\right).
  \label{ps_phia}
  \end{eqnarray}
  Here the mass in the transformed Dirac system is equal to
  $\lambda_2$.

As a conclusion, we have shown that for a pseudoscalar
potential both the transformation operator and the
  transformed potential are expressed in terms of just one function.

\subsection{Interrelation between Darboux transformations for the Dirac and
Schr\"odinger equations}

 It is well known (see e.g. \cite{pra-1998-v57-p93}) that, when using a pseudoscalar
potential, the Dirac system  may be reduced to the following
 two independent Schr\"odinger equations
 \begin{eqnarray}  \label{shred}
 - \psi''_1 + U_0^{(+)} \psi_1  &  = &  \varepsilon \psi_1\,, \\  [1ex]
 - \psi''_2 + U_0^{(-)} \psi_2  &  =  &  \varepsilon \psi_2\,, \label{shredshred}
 \end{eqnarray}
 where
  \begin{equation}\label{U_0pm}
U_0^{(\pm)} = q_0^2 \pm q_0'
  \end{equation}
  and
$
\varepsilon = E^2 - m^2\,.
$
 The system (\ref{shred})--(\ref{shredshred}) is precisely a pair of supersymmetric
Schr\"odinger equations (see e.g. \cite{Bagchi}), one equation being the SUSY partner
of the other. Similarly, one of the Schr\"odinger Hamiltonians \
\begin{equation}
 H^{(\pm)} = - \partial_{xx} + U_0^{({\pm})}
\end{equation}
  is the SUSY partner of the other. Moreover, according to (\ref{ps_v0})  one has
\begin{equation}\label{darbu_vsh}
 U_0^{(-)}=U_0^{(+)}-2(\ln u_{11})'' ,
\end{equation}
where $u_{11}$ is an eigenfunction of $H^{(+)}$ which is everywhere
non-vanishing.
The transformed Dirac equation is also an equation with a
 pseudoscalar potential. Therefore, it can also be reduced to the pair of
 supersymmetric Schr\"odinger equations
 \begin{eqnarray}  \label{sh_phi2}
 - \varphi''_1 + U^{(+)}_1 \varphi_1 & = & \varepsilon_1 \varphi_1\,,
  \\ [1ex]
\label{sh_phi1}
 - \varphi''_2 + U^{(-)}_1 \varphi_2 & = & \varepsilon_1 \varphi_2\,,
  \end{eqnarray}
where
 \begin{equation}
U^{(\pm)}_1 = q_1^2 \pm q'_1\,,
 \end{equation}
$\varepsilon_1 = E^2 - \lambda_2^2$, and $q_1$ is given by
(\ref{ps_v1}).
We see that to this system corresponds an energy different from the one that appears
in equations (\ref{shred})--(\ref{shredshred}).
To compare this system with (\ref{sh_phi2})--(\ref{sh_phi1}),
 we have to displace the energy $\varepsilon_1$.
  For this purpose we add to the left and right hand sides of the equations
  (\ref{sh_phi2}) and (\ref{sh_phi1}) the terms $(\lambda_2^2 - m^2) \varphi_1$ and
  $(\lambda_2^2 - m^2) \varphi_2$,  respectively.
  This leads to shifting the potentials
$U^{(\pm)}_1\to \widetilde{U}^{(\pm)}_1 = U^{(\pm)}_1 + \lambda_2^2 - m^2$.
   Now, taking into account that $u_{12}$ and $u_{22}$ are
   solutions of the system (\ref{shred})--(\ref{shredshred}) with the potentials
  (\ref{U_0pm}) and the expression (\ref{ps_v1}) for the potential  $q_1$, we get
  \begin{eqnarray}  \label{U_1m}
  \widetilde{U}^{(-)}_1 & = & -2 (\ln u_{22})'' + q_0^2 - q'_0=
  \widetilde{U}^{(+)}_1 -2(\ln u_{22})''\,,
   \\  [1ex]
  \widetilde{U}^{(+)}_1 & = & q_0^2 - q'_0 = U^{(-)}_0\,.
  \label{U_1pl}
  \end{eqnarray}
We observe that the potentials  $\widetilde{U}^{(+)}_1$ and $U^{(-)}_0$ coincide and, as
it has been  mentioned in the preceding Section, the functions $\varphi_1$ and
$\psi_2$ satisfy the same equation.

 Using the expression (\ref{ps_v0}) for the potential $q_0$ we
 obtain the potential differences
 \begin{eqnarray}  \label{dU_m}
 \Delta U^{(-)} & = & \widetilde{U}^{(-)}_1 - U^{(-)}_0 =  -2 (\ln u_{22})'',
\\  [1ex]
 \Delta U^{(+)} & = & \widetilde{U}^{(+)}_1 - U^{(+)}_0 =  -2 (\ln u_{11})'',
 \label{dU_pl}
 \end{eqnarray}
 that agree with  (\ref{darbu_vsh}).
 Hence, we can obtain the potential $\widetilde{U}^{(-)}_1$ by two different,
 but equivalent, ways:
\begin{itemize}
\item
 The first possibility is to start with  the initial Dirac system,
 realize the Darboux transformation with the transformation function
 ${\mathcal U}$ given by
 (\ref{uPS}), and then split the resulting Dirac equation
 into a system of two Schr\"odinger  equations,
 related to each other by another Darboux transformation.
 This corresponds to the path
 $q_0\to q_1 \to\widetilde{U}^{(-)}_1$ in the diagram of Figure~1.
\begin{figure}[h]\centering
\epsfig{file=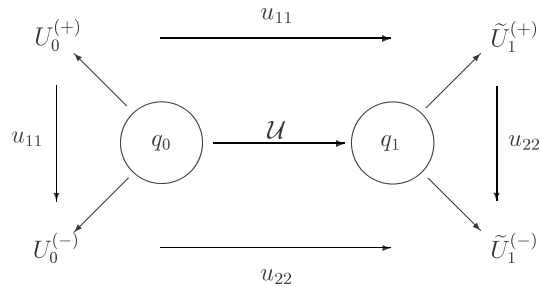, width=70mm}
\begin{minipage}{8cm}
\caption{Diagram showing the
connections between the Dirac system, its transformed, and the
associated SUSY Schr\"odinger equations.}
\end{minipage}
 \label{diagr}
\end{figure}

\item
 The second possibility is to split the Dirac equation into two
 Schr\"odinger equations with potentials $U_0^\pm$, and
 then realize a chain of two transformations at the level of the
 Schr\"odinger equations, starting with the potential $U_0^+$; for the
 first transformation we use $u_{11}$ as the transformation function,
 and get the potential $U_0^-$; finally transforming this potential
 with the help of the transformation function $u_{22}$, we obtain the same
 potential $\widetilde{U}^{(-)}_1$. This path corresponds to
 $q_0\to {U}^{(+)}_0\to {U}^{(-)}_0 \to \widetilde{U}^{(-)}_1$ in the diagram.
\end{itemize}
Other possible path is $q_0\to {U}^{(+)}_0\to
\widetilde{U}^{(+)}_1 \to \widetilde{U}^{(-)}_1$, which  is
completely equivalent to the previous ones, as it is clear from
the preceding discussion. Moreover, as it follows from the scheme
of Figure 1, the following proposition holds.

\begin{proposition}
A Darboux transformation for the Dirac equation with a pseudoscalar potential
{\rm(\ref{pseudo})}, with transformation function {\rm(\ref{uPS})}, is equivalent to a
two-step SUSY transformation for the Schr\"odinger equation with the potential
$U_0^+$ given by {\rm (\ref{U_0pm})} with transformation functions
$u_{11}$ and $u_{22}$.
\end{proposition}

   For the case  $\lambda_1 = -m$, $u_{11} = 0$, similar calculations give us
 \begin{eqnarray}
 \widetilde{U}^{(-)}_1 & = & q_0^2 + q'_0 = U^{(+)}_0,
 \label{UU_1m} \\ [1ex]
 \widetilde{U}^{(+)}_1 & = & -2 (\ln u_{12})'' + q_0^2 + q'_0.
 \label{UU_1pl}
 \end{eqnarray}

 We conclude this Section by stressing that Darboux
 transformation for the Dirac equation induces Darboux
 transformations for the associated Schr\"odinger equations.

\section{Scalar potentials}

\subsection{Darboux transformation for a scalar potential}

In this Section, let us consider that  the initial potential $V_0$ has the scalar form
\begin{equation}\label{ScPot}
V_0(x)=p_0(x)\sigma_3\,,\quad p_0=m+S_0(x) \,,
\end{equation}
where $m$ is the mass of a particle and the function $S_0(x)$ is supposed to be known.
Note first that if the function $\psi $ is a solution of the Dirac
equation with the potential (\ref{ScPot}) and the energy $E$, then
the function $\sigma_1\psi$ is also a solution, but with the energy
$-E$. In particular, this means that the spectrum of the Dirac
equation is symmetric with respect to $E=0$.

In the literature another representation of scalar potentials is frequently used:
\begin{equation} \label{Vpseudo}
\widehat {V}_0(x) = (m + S_0(x)) \sigma_1\ .
\end{equation}
Both representations (\ref{ScPot}) and (\ref{Vpseudo})
 are related by a  unitary
transformation $\widehat U$:
\begin{equation}   \label{trans}
\widehat{V}_0  = \widehat U^{-1} V_0 \widehat U\,, \qquad \widehat{\psi}  =
\widehat U
\psi \ ,
\end{equation}
where
\begin{equation} \label{Umatr}
\widehat U = (1 + \gamma)/ \sqrt{2}\, .
\end{equation}
Observe that the scalar potential (\ref{Vpseudo})
can be considered as a
pseudoscalar potential  (\ref{pseudo}) when the  value of the mass is equal to zero.
Nevertheless, it has some new properties to be considered below.

 The Darboux transformation, in its general form, does not preserve the scalar
 character of a potential. Therefore, it is necessary to select those
 transformations such that they will preserve the scalar form of a potential.
As it follows from (\ref{V1D})--(\ref{d-matr}), the transformed
potential remains to be scalar if $d_2(x)\equiv 0$. From
(\ref{d12}) one easily sees that to satisfy this condition it is
sufficiently to construct the transformation function ${\mathcal U} =(u_1,u_2)$
from the spinors $u_1=(u_{11},u_{21})^t$ and $u_2 = \sigma_1 u_1$ with
eigenvalues
$\lambda_1 = \lambda^{(0)}$ and $\lambda_2 = -\lambda^{(0)}$, respectively.

In order to look for new properties of the Darboux transformation for the
Dirac equation with a scalar potential, it is more convenient to
consider the scalar potentials written in the form (\ref{Vpseudo}). Under the
transformation (\ref{Umatr}) the potential $V_0$ goes into
$\widetilde{V}_0 = (m + S_0) \sigma_1$, and the transformation function
${\mathcal U}$ into $\widehat {\mathcal U}$, given by
\begin{equation}
\widehat{{\mathcal U}} = \left( \begin{array}{cc}
\widehat{u}_{11} & -\widehat{u}_{11} \\
\widehat{u}_{21} & \widehat{u}_{21}
\end{array} \right), \label{tilde_u}
\end{equation}
where
$\widehat{u}_{11} = u_{11} - u_{21}$, $\widehat{u}_{21} = u_{11}+ u_{21}$.
For calculating the new potential, we apply (\ref{V1D})--(\ref{d12}), and
for getting the transformation  operator we use    (\ref{eql}),
where we have to make the replacement ${\mathcal U} \to \widehat{\mathcal U}$.
The matrix $\widehat{{\mathcal U}}_x\, \widehat{{\mathcal U}}^{-1}$ is now diagonal:
\begin{equation}
\widehat{{\mathcal U}}_x\, \widehat{{\mathcal U}}^{-1}  = \left( \begin{array}{cc}
(\ln \widehat{u}_{11})' & 0 \\  [1ex]
0 & (\ln \widehat{u}_{21})'
\end{array} \right).
\label{uu-1}
\end{equation}
From Eq. (\ref{V1}) we find that the transformed potential is
\begin{equation}  \label{v1_tilde}
\widehat{V}_1 = ( m + S_1) \sigma_1,
\end{equation}
where
\begin{equation}\label{S1}
S_1 = S_0 + (\ln \widehat{u}_{21})' - (\ln \widehat{u}_{11})'.
\end{equation}
Solutions of the Dirac equations with the potential
 (\ref{v1_tilde})--(\ref{S1}) are found by the action of the
 operator $L=\partial_x-\widehat{\mathcal U}_x\, \widehat{\mathcal U}^{-1}$  on
solutions of the initial
 equation:
\begin{equation} \label{phi_tilde}
\widehat{\varphi} =  L \widehat{\psi} = \left(
\begin{array}{c}
\widehat{\psi}'_1 - (\ln \widehat{u}_{11})' \widehat{\psi}_1 \\ [1ex]
\widehat{\psi}'_2 - (\ln \widehat{u}_{21})' \widehat{\psi}_2
\end{array}
\right).
\end{equation}

Remark that if in the positive part of the spectrum of the initial
Hamiltonian there exists a ground state level $E=E_0$ with wave
function $\psi_0(x)$, then in the negative part of it there exists a
highest energy level $E=-E_0$ with wave function
$\sigma_1\psi_0(x)$. The choice of the spinors $u_1=\psi_0$ and
$u_2=\sigma_1\psi_0$ generates a potential $V_1$ with the same
spectrum as $V_0$, except for the levels $\pm E_0$.

The use of formula  (\ref{tilde_u})
gives us the matrix solution of the transformed Dirac equation with
the matrix eigenvalue $\Lambda$
\begin{equation} \label{tilde_u-1}
 \widehat{\mathcal V} = (\widehat{{\mathcal U}}^+)^{-1} = \frac{1}{2}
\left( \begin{array}{cc}
(\widehat{u}_{11}^*)^{-1} & -(\widehat{u}_{11}^*)^{-1} \\ [1ex]
(\widehat{u}_{21}^*)^{-1} & (\widehat{u}_{21}^*)^{-1}
\end{array} \right) \equiv (\widehat v_1,\widehat v_2)
\end{equation}
We observe here that the spinors $\widehat v_1$ and $\widehat v_2$ are
either square integrable or non-integrable simultaneously. In the
first case the levels $E=\pm\lambda^{(0)} $ appear in the spectrum
of $h_1$. Hence, the Darboux transformation may create the energy
levels only by pairs symmetrically disposed with respect to $E=0$. This
agrees with the fact that it produces a scalar potential which may
have only a symmetrical spectrum.

   \subsection{Interrelation with the Schr\"odinger equation}

   It is well-known (see e.g. \cite{Nogami-93}) that the Dirac
   system with the scalar potential  (\ref{Vpseudo}) may be reduced to
 the following   supersymmetric pair of the Schr\"odinger equations:
 \begin{equation}
 \begin{array}{lcr}
 - \widehat{\psi}''_1 + U^{(+)} \widehat{\psi}_1 & =
 &\varepsilon \widehat{\psi}_1\,, \\ [1ex]
 - \widehat{\psi}''_2 + U^{(-)} \widehat{\psi}_2 & =
 &\varepsilon \widehat{\psi}_2\, .
 \end{array}
 \label{shredt}
 \end{equation}
 Since the transformed Dirac system also corresponds to a scalar
 potential, it may be associated with a couple of similar
 equations
 \begin{equation}
 \begin{array}{lcr}
 - \widehat{\varphi}''_1 + U^{(+)}_1 \widehat{\varphi}_1 & = &
 \varepsilon_1 \widehat{\varphi}_1\,,\\  [1ex]
 - \widehat{\varphi}''_2 + U^{(-)}_1 \widehat{\varphi}_2 & = &
 \varepsilon_1 \widehat{\varphi}_2\,,
 \end{array}
 \label{sh_phi}
 \end{equation}
where
 \begin{equation}
 U^{(\pm)}_1 = (m + S_1)^2 \pm S'_1.
 \label{u1pm}
 \end{equation}
 Taking into account the equation for $S_1$  (\ref{S1}),
 after some algebra
 we get from (\ref{u1pm}) the potentials $U^{(\pm)}_1 $
 \begin{eqnarray}
 U^{(+)}_1 & = & (m + S_0)^2 + S_0' - 2 (\ln \widehat{u}_{11})''\, ,
 \label{u1pl}\\   [1ex]
 U^{(-)}_1 & = & (m + S_0)^2 - S_0' - 2 (\ln \widehat{u}_{21})''\, .
 \label{u1mi}
 \end{eqnarray}
Hence, we also conclude that the potentials $U^{(\pm )}_1$ are SUSY
 partners of the potentials $U^{(\pm )}_0$, and Darboux
 transformation for the Dirac equation induces Darboux
 transformations of corresponding supersymmetric pair of
 initial Schr\"odinger equations.

\section{Illustrative examples}

In this Section we will show how the technique we have developed for Dirac systems is
applied to some interesting examples of pseudoscalar and scalar potentials, as well as
to spherically symmetric potentials.

\subsection{Pseudoscalar potentials}

First, we will analyse some examples of Darboux transformation applied to transparent
potentials, and also to the relativistic harmonic oscillator.

\subsubsection{Transparent potentials}

In this paper we do not analyse the changes in transmission  and reflection
coefficients produced by Darboux transformations.
Nevertheless, we can easily notice that
 if the initial potential is transparent, i.e., if it produces the zero reflection
 coefficient for an incident particle, or if the potential
 does not change  the  asymptotic
 form of a continuous spectrum eigenfunction, then the transformed
 potential keeps this property unchanged. In particular, this means that starting
 with the free particle Dirac equation we shall get only transparent potentials.
Hence, in this Section let us consider the potential
\begin{equation}   \label{transparent1}
V_0 = m \sigma_3,
\end{equation}
which is a particular case of (\ref{pseudo}) with $q_0=0$.

\begin{example}{\rm
Let us take as transformation function the following matrix
\begin{equation}   \label{u0}
{\mathcal U} = \left(\begin{array}{cc}
1 & \frac{k}{\varepsilon - m} \ \sinh kx \\ [1ex]
0 & \cosh kx
\end{array}\right)\,,
\quad k = \sqrt{m^2-\varepsilon^2}\,.
\end{equation}
 It corresponds to
 $\lambda_1 = m$, $\lambda_2 = \varepsilon <  m$.
 The transformed potential is obtained by Eqs. (\ref{ps_V1})
 and (\ref{ps_v1}), that generate the well-known one-soliton
potential (see e.g. \cite{NogToy98,Thaler}):
 \begin{equation}  \label{v1_th}
 V_1 = - \varepsilon \sigma_3 + k \tanh kx \, \sigma_1 \ .
 \end{equation}
From Eq. (\ref{u0}) we find the matrix solution of the transformed
  Dirac equation for the particular value $E=\Lambda=\mbox{diag}(m,\varepsilon)$:
\begin{equation}  \label{v1_uplus}
{\mathcal V}=({\mathcal U}^+)^{-1} = \left(\begin{array}{cc}
1 & 0 \\  [1ex]
- \frac{k}{\varepsilon - m}\, \tanh k x & \mbox{sech\,} kx
\end{array}\right)   \,.
 \end{equation}
 We conclude from (\ref{v1_uplus}) that the potential (\ref{v1_th})
  has one discrete level at $E=\varepsilon$.
  The solutions of the Dirac equation with the potential
  (\ref{v1_th}) at $E\ne \varepsilon , m$ may be found by applying the
 operator $L$ to solutions $\psi$ of the free particle equation:
 \begin{equation}  \label{newv}
 {\varphi } = L \psi = \psi_x - {\mathcal U}_x\,  {\mathcal U}^{-1} \psi \, .
 \end{equation}

The potential (\ref{v1_th}) may be considered as the initial potential for the
 next transformation step. To carry it out, we need a spinor solution of the
 Dirac equation with either upper or lower component equal to
 zero. If we take as a solution of the free particle Dirac equation
 \begin{equation}  \label{solutionfreediraceq}
 \psi_1 =
\left(\begin{array}{c}
-\frac{k}{\varepsilon+m}\ \sinh kx \\  [1ex]
\cosh kx
\end{array}\right)
\end{equation}
 then, from (\ref{newv}) we find the following spinor solution of the Dirac
equation for the potential (\ref{v1_th}):
 \begin{equation}
{\varphi }_1 =
\left(\begin{array}{c}
-2\varepsilon \cosh kx\\  [1ex]
0
\end{array}\right)
 \end{equation}
 corresponding to the eigenvalue $\lambda_1 =-\varepsilon.$
 Another solution
 of the Dirac equation for this potential (\ref{v1_th})
 may be found with the function
 \begin{equation}
\psi_2 =
\left(\begin{array}{c}
-\frac{k_1}{\varepsilon_1-m}\ e^{-k_1x} \\ [1ex]
 e^{-k_1x}
\end{array}\right)
 \end{equation}
 in (\ref{newv}), which gives us
 \begin{equation}
{\varphi }_2 =
\left(\begin{array}{c}
(\varepsilon - \varepsilon_1) e^{-k_1x} \\ [1ex]
  (k_1 +k \tanh kx) e^{-k_1x}
\end{array}\right)   , \quad
k_1 = \sqrt{m^2 - \varepsilon_1^2}.
 \end{equation}
This spinor has the eigenvalue $\lambda_2 =\varepsilon_1$.
From the spinors $ u_1=\varphi _1$ and $u_2=\varphi _2$ we construct the matrix
 solution for the potential (\ref{v1_th}), ${\mathcal U}=(u_1,  u_2)$,
 with the matrix eigenvalue $\Lambda =\mbox{diag}(-\varepsilon ,\varepsilon_1)$.
}
\end{example}

\begin{example}{\rm
 When the above matrix solution ${\mathcal U}$ is taken as the transformation
function for the second  transformation step, it  produces a two-soliton 
potential of the form:
  \begin{equation}   \label{v2b}
 V_2 = -\varepsilon_1 \sigma_3 +
  \left(\frac{k^2 - k_1^2}{k_1 + k \tanh kx} - k \tanh kx \right) \sigma_1 \ .
  \end{equation}
  The choice $k_1 > k>0$ assures the regular behaviour of this
  potential $V_2$, which keeps the discrete level $E=\varepsilon$ unchanged and has
an additional level at $E=-\varepsilon$.
The last statement can be easily seen from
the matrix solution for the potential  (\ref{v2b}) with matrix eigenvalue
   $\Lambda =\mbox{diag}(-\varepsilon ,\varepsilon_1)$:
   $$
   {\mathcal V}=({\mathcal U}^+)^{-1}  = \left( \begin{array}{cc}
   \frac{1}{2 \varepsilon \cosh kx} & 0 \\  [2ex]
   \frac{\varepsilon - \varepsilon_1}{2 \varepsilon \cosh kx(k_1 + k \tanh
kx)} &
   \frac{-e^{k_1x}}{k_1 + k \tanh kx}
   \end{array} \right).
   $$
The first column  of this matrix is a square integrable spinor with eigenvalue
$E=-\varepsilon$.
}
\end{example}

\begin{example}{\rm
In (\ref{newv}) let us choose now
\begin{equation}
\psi_2 =
\left(\begin{array}{c}
\frac{k_1}{\varepsilon_1-m}\ \cosh k_1x  \\ [1ex]
\sinh k_1x
\end{array}\right)
, \qquad
k_1 = \sqrt{m^2 - \varepsilon_1^2} \,.
\end{equation}
This gives us
\begin{equation}
\varphi _2 =
\left(\begin{array}{c}
(\varepsilon - \varepsilon_1) \sinh k_1x  \\ [1ex]
 k_1 \cosh k_1x  -k \tanh kx\ \sinh k_1x
\end{array}\right)  \,,
\quad k = \sqrt{m^2 - \varepsilon^2} \,.
\end{equation}
The use of this
 spinor
 as the second component of the transformation function,
 $u_2=\varphi_2$,
 together with the previously found $u_1$ as the first component,
 ${\mathcal U}=(u_1,u_2)$,
 produces the following three-soliton potential:
 \begin{equation}  \label{v2c}
V_2 = -\varepsilon_1 \sigma_3 +
 \left(\frac{k_1^2 - k^2}{k_1 \coth k_1x - k \tanh kx} - k \tanh kx \right)
 \sigma_1 \ ,
 \end{equation}
which is regular provided $k_1 > k>0$, and has three discrete
levels:
  $E = \pm \varepsilon$ and $E = \varepsilon_1$.

It is important to stress that similar potentials have been  found recently by other
means \cite{NogToy98}. In contradistinction to these authors,
we are able to indicate
precisely the position of the discrete
  levels.
  Moreover, corresponding wave functions are easily obtained from
  columns of the matrix-function $({\mathcal U}^{+})^{-1}$
  In the next example we give more general
  transparent potential with three discrete levels.
}\end{example}

\begin{example}{\rm
 We can take   the following spinor as a solution of the free particle
  equation in order to get a solution of the Dirac equation
  with the one-soliton potential, which is:
\begin{equation}
\psi_2 =
\left(\begin{array}{c}
\frac{k_1}{\varepsilon_1-m}(\sinh k_1x + B \cosh
  k_1x) \\  [1ex]
(\cosh k_1x + B \sinh k_1x)
\end{array}\right)  ,
\end{equation}
  where $B$ is an arbitrary constant. From (\ref{newv}), we obtain
\begin{equation}
 \varphi_2  =
\left(\begin{array}{c}
(\varepsilon - \varepsilon_1)
 Q(x) \\  [1ex]
 k_1(\sinh k_1x + B \cosh k_1x) - k Q(x)\tanh kx )
\end{array}\right)  ,
\end{equation}
 where $Q(x)=\cosh k_1x + B \sinh k_1x$. When $\varphi_2$ is used as the spinor
$u_2$ for the second transformation step, it produces the potential
 \begin{eqnarray}   \label{v2a}
   V_2  &=&  -\varepsilon_1 \sigma_3 \\
&& +
 \left(\frac{(k_1^2 - k^2)(1 + B \tanh k_1x)}
 {k_1 (\tanh k_1x + B) - k \tanh kx (1 + B \tanh k_1x)} - k \tanh kx \right)
 \sigma_1 \, .  \nonumber
 \end{eqnarray}
 It is not difficult to prove that it is regular provided
 $k_1 > k>0$ and $B > 1$, and also that it has three discrete levels:
 $E = \pm \varepsilon$ and $E = \varepsilon_1$.
 A typical behavior of this potential is shown in Figure~2, where the
term between parentheses in Eq.~(\ref{v2a}) is plotted. One can notice
that closer is the parameter
$B$ to the value
$1$, wider is the potential barrier.
\begin{figure}[htbp]\label{figex4}
 \centering
\epsfig{file=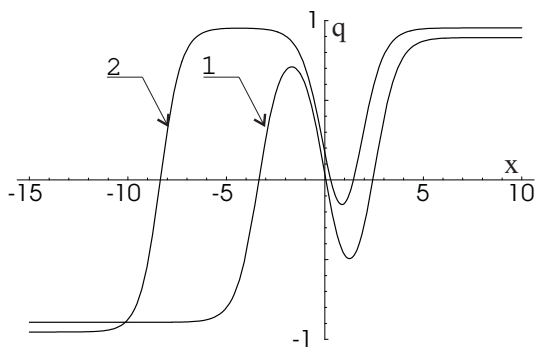, width=70mm}
\begin{minipage}{8cm}
\caption{Three-level transparent pseudoscalar
potentials with
$m=1$ and $\varepsilon =0.5$.
Curve 1 is obtained for $B=1.5$ and $\varepsilon =0.45$
Curve 2 is obtained for $B=1.000005$ and $\varepsilon =0.3$.}
\end{minipage}
\end{figure}
 A potential of this type is not known in the available literature.
}
\end{example}

 To finish this subsection, we would like to remark that our method can produce
transparent potentials of a more general form, which are a superposition of scalar and
pseudoscalar potentials.  We illustrate this fact in he next example

\begin{example}{\rm
 As a final case of transparent potentials, let us consider as transformation function the
following matrix solution of the free particle Dirac equation:
 \begin{equation}
 {\mathcal U} =  \left(\begin{array}{cc}
 -\frac{k}{\varepsilon + m}\ \sinh kx  & \frac{k}{\varepsilon - m}\ e^{kx} \\ [1ex]
 \cosh kx & e^{kx}
 \end{array}\right),\ 
  \lambda_1 = -\varepsilon ,\   \lambda_2 = \varepsilon ,
  \  k=\sqrt{m^2-\varepsilon^2}.
 \label{u0a}
 \end{equation}\
 It generates a completely new potential of the form
 \begin{equation}
 V_1 = \left[
 m-\frac{2k^2}{\varepsilon {e^{2kx}}^{\vphantom j}+m}
 \right]\sigma_3 +
 2 k \varepsilon\,
 \frac{\sinh kx - \cosh kx}{\varepsilon e^{kx}+m {e^{-kx}}^{\vphantom j}}
 \ \sigma_1 \,.
  \label{v1a}
 \end{equation}
 From the analysis of the function
$$
({\mathcal U}^+)^{-1} = \left(
\begin{array}{cc}
\displaystyle\frac{k }{\varepsilon e^{kx}+m {e^{-kx}}^{\vphantom j}}    &
\displaystyle\frac{-k\cosh kx}{\varepsilon {e^{2kx}}^{\vphantom j}+m}
\\  [2ex]
\displaystyle\frac{m+\varepsilon }{\varepsilon e^{kx}+m {e^{-kx}}^{\vphantom j}}  &
\displaystyle\frac{(\varepsilon  -m)\sinh kx}{\varepsilon {e^{2kx}}^{\vphantom j}+m}
\end{array}
\right)
$$
we conclude that $V_1$ has one discrete level $E = -\varepsilon$.
}
\end{example}

\subsubsection{Darboux transformation for the Dirac oscillator}

As it is well known, in the literature there are several possible candidates to be
considered as the relativistic Dirac oscillator. We choose the one which is described by
the Hamiltonian
\cite{Toyama-97}
\begin{equation}
  h_0 = \gamma \partial_x + m \sigma_3 + \frac{x}{2} \sigma_1\,.
  \label{d_osc}
  \end{equation}
Its discrete spectrum consists of a positive series
  \begin{equation}
E^{(+)}_n = (m^2 + n)^{1/2},  \qquad n = 1,2,\ldots
  \end{equation}
 and a negative series
  \begin{equation}
E^{(-)}_n = -(m^2 + n)^{1/2}, \qquad n = 0,1,\ldots.
  \end{equation}
We will use this model to illustrate the aplications of the Darboux transformation for
the Dirac equation in three examples.

\begin{example}{\rm
 We will use the spinors
\begin{equation}
u_1 =
\left(\begin{array}{c}
  e^{x^2/2}   \\  [1ex]   0  \end{array} \right),
 \label{phi_m}
  \end{equation}
and
\begin{equation}
 u_2=
\left(\begin{array}{c}
-\frac{i}{\varepsilon_n - m}\ e^{x^2/4}\ \mbox{He}_n(ix)  \\ [2ex]
  e^{x^2/4}\ \mbox{He}_{n-1}(ix)
\end{array} \right) ,
  \label{phi_en}
 \end{equation}
 for constructing the transformation function ${\mathcal U} =(u_1,u_2)$. The
 spinor (\ref{phi_m}) is a solution of the Dirac equation with the
 potential (\ref{d_osc}) for $\lambda_1 = m$; the same is true for the spinor
 (\ref{phi_en}) with eigenvalue  $\lambda_2 = \varepsilon_n = (m^2 - n)^{1/2}$.
  Here
\begin{equation}
\mbox{He}_n(z) = 2^{-n/2} H_n(z/\sqrt{2})\,,
\end{equation}
where $H_n(z)$ are Hermite polynomials.
Using the formulas (\ref{ps_V1}) and (\ref{ps_v1}) we obtain the
transformed potential
\begin{equation} \label{v1_ex1}
V_1 = -\varepsilon_n \sigma_3 + \left(\frac{x}{2} + (n-1)\
\frac{K_{n-2}(x)}{K_{n-1}(x)} \right) \sigma_1\,,
\end{equation}
where $K_n(x) = (-i)^n\, \mbox{He}_n(ix)$. Observe that the functions
$K_n(x)$ are real and nodeless for even values of $n$; for odd
values of $n$ they have only one node at $x=0$ \cite{10}.
Therefore, the potentials (\ref{v1_ex1}) are real and regular when $n$ takes odd
values. The analysis of the function $({{\mathcal U}^+})^{-1}$ shows that this
 potential has two additional discrete levels
 $E = m$ and $E = \varepsilon_n$ with respect to the initial
 harmonic oscillator potential (\ref{d_osc}).
}\end{example}

\begin{example}{\rm
Let us keep the spinor $u_2$  as in the previous example and let us
 take $u_1  =  (0, \ e^{-x^2/2})^t$, which corresponds to $\lambda_1 = -m$.
Using the same procedure of the previous example, we get the potential
 \begin{equation}
 V_1 = \varepsilon_n\, \sigma_3 -
 \left(\frac{x}{2} + n\ \frac{K_{n-1}(x)}{K_n(x)} \right) \sigma_1\,  .
 \label{v1_ex2}
 \end{equation}
 In contrast to the previous example, $\lambda_1$ belongs now to
 the discrete spectrum of the initial potential. Therefore, the
 level  $E = -m$ is deleted from the spectrum of the
 Hamiltonian (\ref{d_osc}) and the new level $E = \varepsilon_n$ is
 added.
 The potential (\ref{v1_ex2}) is everywhere  regular for even
 values of $n$.
}\end{example}

\begin{example}{\rm
 Finally, let us take now the spinor $u_1$ as in Example 1 and choose the
 spinor $u_2$ as follows:
 \begin{equation}
 u_2 =
\left(\begin{array}{c}
\frac{1}{\lambda_2-m}\
 (x\, Q(x)\ e^{x^2/4} + e^{-x^2/4})  \\  [1ex]
 Q(x)\ e^{x^2/4}
\end{array} \right) ,
  \label{phi_erf}
 \end{equation}
which corresponds to $\lambda_2 = (m^2 - 1)^{1/2}$.
 Here, $Q(x) = \sqrt{\pi/2}\, (B + \mbox{erf}(x/\sqrt{2}))$, with $|B| > 
1$.
 We get for the transformed potential
 \begin{equation}
 V_1 = -\lambda_2 \sigma_3 + \left( \frac{x}{2} +
 e^{-x^2/2}\, Q^{-1}(x) \right) \sigma_1\,.
 \label{v1_erf}
 \end{equation}
 A simple analysis shows that this potential has two additional
 discrete levels $E = m$ and $E = (m^2 - 1)^{1/2}$ with respect to
 the initial potential (\ref{d_osc}).
 Two typical representatives of this potential are displayed
 in Figure~3, where the term between parentheses in Eq.~(\ref{v1_erf}) is
plotted. It is clearly seen that closer is the parameter $B$ to the value
$1$, larger is the perturbation of the initial potential.
 \begin{figure}[htbp]  \label{figex8}
 \centering
\epsfig{file=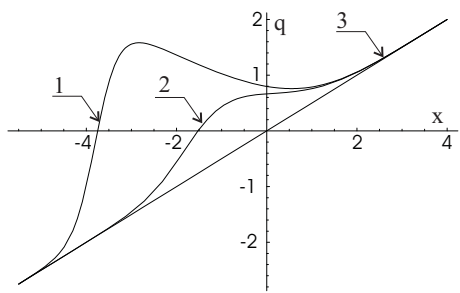, width=60mm}
\begin{minipage}{8cm}
\caption{Harmonic oscillator potential $q_0(x)=x/2$ (curve~3) with $m=1$,
and two of its Darboux transformed partners $q_1(x)$. Curve~1 is obtained
for $B=1.0002$, and  curve~2 for $B=1.2$.}
\end{minipage}
\end{figure}
}
\end{example}

\subsection{Scalar potentials}

In this subsection we will analise some scalar potentials of the form (\ref{Vpseudo})
$$
\widehat {V}_0(x) =p_0(x)\sigma_1 = (m + S_0(x)) \sigma_1\,.
$$
In the first instance, we will consider the simplest case, corresponding to
$S_0(x)\equiv 0$.

\subsubsection{Transparent potentials}

\begin{example}{\rm
The spinor
\begin{equation}
\widehat{u}_1  =
\left(\begin{array}{c}
e^{kx} + e^{-kx}  \\  [1ex]
\frac{m-k}{\lambda}\ e^{kx} +
\frac{m+k}{\lambda}\ e^{-kx}
\end{array} \right) ,  \qquad
k = \sqrt{m^2 - \lambda^2},
\label{ex1-u}
\end{equation}
is a solution of the Dirac equation with potential  $\widehat{V}_0 = m \sigma_1$  for
$E= \lambda < m$. Choosing the transformation function in the form
$\widehat {\mathcal U}=(\widehat u_1,\widehat  u_2)$,
$\widehat u_2=-\sigma_3\widehat u_1$, and using Eq.~(\ref{S1}), we get the
transformed potential
\begin{equation}
S_1  = -\frac{2k^2}{m + \lambda \cosh(2kx+2\alpha)},
\label{ex1-s1}
\end{equation}
where  $e^{2\alpha} = \sqrt{(m-k)/(m+k)}$.
This is a transparent potential, which has been previously found in \cite{Nogami-93}.
It is easy to see that the two spinors coming from the matrix
$(\widehat{{\mathcal U}}^{+})^{-1}$ are square integrable. This means that the
potential (\ref{ex1-s1}) has two discrete levels: $E = \pm \lambda$.
}
\end{example}

\begin{example}{\rm
Consider now the spinor
\begin{equation}
\widehat{v}_1  =
\left(\begin{array}{c}
\sinh k_1 x\\  [1ex]
\sinh (k_1 x + 2 \alpha_1)
\end{array} \right) ,  \quad
k_1 = \sqrt{m^2 - \lambda_1^2}, \ \ e^{2\alpha_1} = \sqrt{\frac{m-k_1}{m+k_1}},
\label{ex2-v}
\end{equation}
satisfying the free Dirac equation for $E =\lambda_1<m$. After acting on it with the
transformation operator of the Example 1, we get the spinor
 \begin{equation}
 \widehat{w}_1  =
\left(\begin{array}{c}
 k_1 \cosh k_1 x - k \tanh kx\, \sinh k_1 x  \\  [1ex]
 k_1 \cosh (k_1 x + 2\alpha_1) - k \tanh kx\, \sinh (k_1 x + 2\alpha_1)
\end{array} \right),
\label{ex2-w}
 \end{equation}
 which is a solution of the Dirac equation with the potential
 (\ref{ex1-s1}). If we choose the spinors
 $\widehat{w}_1$ and
 $\widehat{w}_2 = -\sigma_3 \widehat{w}_1$ for the next
 transformation step, we obtain a two-step potential
 \begin{eqnarray}
 S_2  &=& \frac{k_1^2 - k^2}{k_1 \coth (k_1 x + 2\alpha_1)  -
 k \tanh(kx+2\alpha)}
\nonumber 
\\ &&
- \frac{k_1^2 - k^2}{k_1 \coth k_1 x - k \tanh kx}\,.
 \label{ex2-s2}
 \end{eqnarray}
 For  $0<k< k_1<m$ this is a regular transparent potential with four
 discrete levels  $E = \pm \lambda$ and  $E = \pm \lambda_1$.
 Figure~4 shows the typical shape of such potentials.
From this figure it is clearly seen that closer are the discrete levels of
the potential, more distant from each other are the two potential wells.
  \begin{figure}[htbp] \label{figex10}
  \centering
\epsfig{file=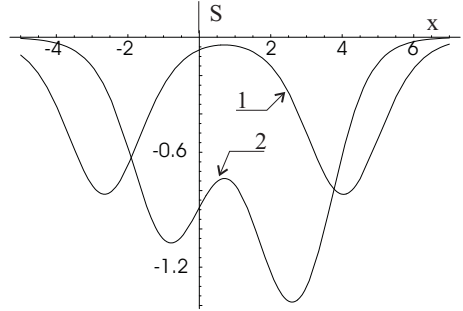, width=60mm}
\begin{minipage}{8cm}
\caption{Four-level scalar potentials with $m=1$ and
$\lambda =0.6$. Curve 1 is obtained for $\lambda_1=0.58$, and  curve 2
for $\lambda_1=0.2$.}
\end{minipage}
\end{figure}
}
\end{example}

\subsubsection{Scalar Coulomb potential}

We will analise now radial potentials of the form
 \begin{equation}
\widehat{V}_0 = \left(m - \frac{\alpha}{r}\right) \sigma_1, \qquad \alpha > 0,
 \end{equation}
which find applications in
modelling inter-quark interactions \cite{Benvegnu}. In this case, the
discrete  spectrum of the
 Dirac Hamiltonian consists of a positive and a negative series  \cite{Dominguez}
\begin{equation}
 E_n = \pm m\  \sqrt{1 - \frac{\alpha^2}{(n + \alpha)^2}}, \quad  n = 1, 2, \ldots,
\end{equation}
plus the zero energy level  $E=0$ \cite{Khalilov}.
The eigenfunctions of the discrete spectrum  (not normalized here) have the
form
 \begin{equation}
 \widetilde{\psi}_n  =
 \left(\begin{array}{c}
 - \frac{E_n\, (n-1)!}{\varepsilon_n (2\alpha +1)_n}\
 e^{-x} \ x^{\alpha+1}\ L^{2\alpha+1}_{n-1}(2x)   \\  [2ex]
 \frac{n!}{(2\alpha)_n}\ e^{-x}\ x^{\alpha}\ L^{2\alpha-1}_n(2x)
\end{array} \right),
 \label{psi-n}
 \end{equation}
 where $\varepsilon_n = \sqrt{m^2-{E_n^2}^{\vphantom{k}}}$
 and $x = \varepsilon_n r$, $n = 1,2,\ldots$ .

\begin{example}{\rm
In order to apply the two-step Darboux transformation
described in Section 7, let us take the spinors $\widehat {\mathcal U}=(\widehat
u_1,\widehat u_2)$, with $\widehat{u}_1 = \widehat{\psi}_k$ and $\widehat{u}_2
= -\sigma_3
\widehat{\psi}_k$, and  $\widehat {\mathcal V}=(\widehat v_1,\widehat v_2)$, with
$\widehat{v}_1 = \widehat{\psi}_{k+1}$ and
$\widehat{v}_2 = -\sigma_3 \widehat{\psi}_{k+1}$.
For the transformed potential we obtain
\begin{equation}
\widehat {V}^{(k)}_2(x) = (m + S^{(k)}_2(x)) \sigma_1, \ 
     S^{(k)}_2  =  - \frac{\alpha}{r} + (\ln{Q_2(r)})' - (\ln{Q_1(r)})' ,
\label{ex3-s2}
\end{equation}
where
\begin{eqnarray}
     Q_1(r)   & = &   \frac{k L^{2\alpha+1}_{k}(2\varepsilon_{k+1}r)
L^{2\alpha-1}_{k}(2\varepsilon_{k}r)}{2\alpha+k+1} -
\frac{(k+1) L^{2\alpha+1}_{k-1}(2\varepsilon_{k}r)
L^{2\alpha-1}_{k+1}(2\varepsilon_{k+1}r)}{2\alpha+k} ,
\nonumber \\  [1ex]
     Q_2(r)   & = &    \frac{L^{2\alpha+1}_{k}(2\varepsilon_{k+1}r)
L^{2\alpha-1}_{k}(2\varepsilon_{k}r)}{(\alpha+k+1)^2} -
\frac{L^{2\alpha+1}_{k-1}(2\varepsilon_{k}r)
L^{2\alpha-1}_{k+1}(2\varepsilon_{k+1}r)}{(\alpha+k)^2}  .
\nonumber
\end{eqnarray}
The spectrum of $\widehat {V}^{(k)}_2(x)$ differs from the spectrum of the initial
Coulomb potential by the absence of the levels  $E = E_k$ and $E = E_{k+1}$.
 We would like to remark that after the first transformation,
 either with the spinor $\widehat u_1$ or with $\widehat u_2$, we get potentials
with singularities, but the second transformation removes all
 singularities and the potential  (\ref{ex3-s2}) is regular for
 $r > 0$.
The simplest particular case corresponds to  $k=1$:
 \begin{equation}
\label{ex3-s2-1}
 S^{(1)}_2 = - \frac{\alpha+2}{r} +
 \frac{2m(2mr - 2\alpha -3)}{2m^2r^2 - 2m(2\alpha+3)r + (\alpha+2)(2\alpha+3)}
 \,.
 \end{equation}
The behavior of the potential (\ref{ex3-s2}) at $k=4$, $m=1$ and
$a=1$ is shown in Figure~5.
 \begin{figure}[htbp]   \label{figex11}
 \centering
\epsfig{file=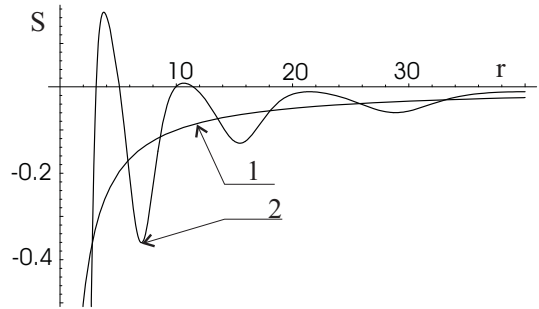, width=70mm}
\begin{minipage}{8.3cm}
\caption{Scalar Coulomb potential $S_0(r)$ (curve~1) and its Darboux
transformed partner $S_2^{(4)}(r)$ (curve~2).}
\end{minipage}
\end{figure}
}\end{example}

\subsection{Spherically symmetric potentials}

The first important detail to be taken into account is that the usual (3+1)-di\-mensional
Dirac equation with a spherically symmetric potential is included in our developments.
 Indeed, using the standard technique of separation of angular
 variables (see e.g. \cite{Weidman}), we get the following
 radial equation coupled to scalar $W_s(r)$, pseudoscalar
 $W_p(r)$ and vector $W_v(r)$ potentials:
\begin{equation}
\left\{\frac{d}{dr}- \left[ \frac{k}{r}+W_P \right] \sigma_3
+[m+W_s ]\sigma_1 + i[E-W_v ]\sigma_2\right\}\psi(r)=0,
\label{eq41}
\end{equation}
where $r$ is the radial variable, $m$ is the mass of the particle, and $E$ its
 energy, while $k=\pm 1,\pm 2 \ldots$ is related to the total
angular momentum. Eq.~(\ref{eq41}) can be also written as
\begin{equation}
\left\{i\sigma_2 \frac{d}{dr}+
\left[ \frac{k}{r}+W_P \right] \sigma_1+\left[m+W_s \right]\sigma_3
-\left[E-W_v \right]  \right\}\psi(r)=0\,,
\label{eq42}
\end{equation}
which if $W_v(r)\equiv 0$ coincides with $h_0 \psi(r)=E \psi(r)$,
$h_0=\gamma\partial_r+V_0(r)$, being $V_0(r)$
given in the canonical form (\ref{CanonV}) with
\begin{equation}
p_0(r)=m +W_s(r)\,,\quad q_0(r)=W_p(r)+\frac kr\,.
\label{eq43}
\end{equation}

Let us start with an unphysical potential, corresponding to
$p_0(r)=q_0(r)=0$ and  $k=0$,  i.e. $V_0=m\sigma_3$.
The associated Dirac Hamiltonian will be denoted by $h_0$ and its eigenfunctions
by $\psi (E)$, where we omit the evident dependence on the variable $r$ and we
stress only the dependence on the eigenvalue $E$. We will show that, starting with
$h_0$, one can obtain physically meaningful potentials with $k\ne 0$. For this purpose,
we shall use the eigenfunctions of $h_0$ with $E=\pm m$:
 \begin{eqnarray}
\psi (m)=
\left(\begin{array}{c}
1 \\ 0
\end{array} \right) \,, &&
\widetilde\psi (m)=
\left(\begin{array}{c}
-2 m r  \\ 1
\end{array} \right)  \,.
 \\ [1ex]
\psi (-m)=
\left(\begin{array}{c}
0 \\ 1
\end{array} \right) \,, &&
 \widetilde\psi (-m)=
\left(\begin{array}{c}
1   \\  -2 m r
\end{array} \right)\,.
 \end{eqnarray}
Other eigenfunctions will  be also used for producing new nontrivial potentials
  \begin{equation}
\psi (\lambda )=
\left(\begin{array}{c}
\cosh kr \\ [1ex]
-\frac{k}{\lambda +m}\ \sinh kr
\end{array} \right) \,,
 \end{equation}
 \begin{equation}
\widetilde\psi (\lambda )= e^{-kr}
\left(\begin{array}{c}
1  \\  [1ex]
\frac{k}{\lambda +m}
\end{array} \right)\,,\quad
k^2=m^2-\lambda ^2>0\,.
 \end{equation}

\begin{example}{\rm
Let us take the spinors $u_1$ and $u_2$ from which the transformation function
${\mathcal U}$ is constructed as follows:
$u_1=\psi (m)$, $u_2=\widetilde\psi (-m)$. After a very simple algebra, we obtain
the potential
$$
V_1=m\sigma_3+\frac 1r \sigma_1,
$$
which is just the free particle Dirac Hamiltonian with $k=1$. It is not difficult to see
that the choice $u_1=\psi (-m)$, $u_2=\widetilde\psi (m)$ gives the same Hamiltonian,
but with $k=-1$.

 Now, all solutions of the Dirac equation with the potential
 $V_1=m\sigma_3+\frac 1r\sigma_1$
 can be found either by applying the transformation operator
 (\ref{eql}) to solutions of the same equation with
 $V_0=m\sigma_3$, or by applying formulas
 (\ref{usm1}) and (\ref{wtpsi1})--(\ref{wtpsi4}).
 Two solutions with $E=m$ are
 \begin{equation}
 \varphi(m)=
\left(\begin{array}{c}
1 \\  \frac{1}{2mr}
\end{array} \right)\,,\quad
  \widetilde\varphi(m)=
\left(\begin{array}{c}
r  \\  0
\end{array} \right).
 \end{equation}
For $E=-m$ one gets
\begin{equation}
\varphi(-m)=
\left(\begin{array}{c}
0 \\  {1}/ {r}
\end{array} \right) \,,\quad  \widetilde\varphi(-m)=
\left(\begin{array}{c}
-3r  \\  2mr^2 \end{array} \right).
\end{equation}
 The solutions with $E\ne \pm m$ have the form
\begin{eqnarray}
&&\varphi(\lambda )=L\psi (\lambda )=
k
\left(\begin{array}{c}
-\sinh kr \\  [1ex]
\frac{-k}{\lambda -m}\ \cosh kr +\frac{1}{(\lambda -m)r} \ \sinh kr
\end{array} \right)\,,\\ [2ex]
&&\widetilde\varphi(\lambda )=L\widetilde\psi (\lambda )=
k e ^{-kr}
\left(\begin{array}{c}
1  \\  \frac{k}{\lambda }+\frac{1}{\lambda r}
\end{array} \right)  \,.
\end{eqnarray}
If we take now the spinors $u_1=\widetilde\varphi(m)$ and
$u_2=\widetilde\varphi(-m)$ for constructing the transformation function
${\mathcal U}$ of the second transformation step, we find after some algebra the
free Dirac Hamiltonian with $k=2$.
}
\end{example}

In the next example we illustrate how the use of a spinor having one of its components
equal to zero inside the transformation function produces a pseudoscalar potential.

\begin{example}{\rm
 The spinors $u_1=\widetilde\varphi(m)$ and $u_2=\widetilde\varphi (\lambda )$
give the simplest pseudoscalar potential with the mass equal to $-\lambda $
 \begin{equation}
q_2(r)= -\frac 1r -\frac{k^2r}{1+kr}\,,
 \end{equation}
  whereas the choice $u_2=\varphi (\lambda )$ corresponds to
  \begin{equation}
 q_2(r)= \frac 2r -
 \frac{3 k r \cosh k r - (3 + k^2 r^2) \sinh kr}
 {r (k r \cosh k r - \sinh k r)}\,.
  \end{equation}
 Another possibility to get a pseudoscalar potential is by choosing
$u_1=\widetilde\varphi (\lambda )$ and  $u_2=\widetilde\varphi (-\lambda )$. By
this means we get the
 potential corresponding to the same mass $m$ and
    \begin{equation}
   q_2(r)= \frac 2r -
   \frac{k\cosh k r + k^2 r^2 \mbox{cosech}\, k r- 2 \sinh k r}
   {r (k r \cosh k r - \sinh k r)}    \,.
    \end{equation}
}\end{example}

 In  the final example we are going to consider in this paper, a superposition of scalar
and pseudoscalar potentials is shown.

\begin{example}{\rm
We use here a linear combination of two spinors with eigenvalue $m$:
 $u_1=c\varphi (m)+\widetilde\varphi (m)$ ($c$ is a constant)
 and $u_2=\widetilde\varphi (-m)$. Then, we obtain a potential with
   \begin{eqnarray}
   p_2(r)&=&2m-\frac{16(c m^3 r^2 + 2 m^4 r^3)}{3 c + 4 c m^2 r^2 + 8 m^3
   r^3}\,,
   \\ [1ex]
  q_2(r)&=&-\frac 1r +
  \frac{8 m^2 r(2 c + 3 m r)}{8 m^3 r^3 + c(3 + 4 m^2 r^2)} \,.
   \end{eqnarray}
This potential is regular provided $c>0$.

Finally, we choose a transformation function composed by following spinors:
$u_1=\varphi (m)$ and $u_2=\varphi (\lambda )$, and we produce another exactly
solvable potential corresponding to
   \begin{eqnarray}
   p_2(r) &=&-\frac{2 k m r \lambda \cosh k r
    +(k^2 + m(m - \lambda ) )\sinh k r}
   {2 k m r \cosh k r + ( \lambda -m) \sinh k r}\,,
   \\ [2ex]
  q_2(r) &=&k\ \frac{2 k m r \sinh k r-k (m + \lambda ) \cosh k r}
  {2 k m r \cosh k r + ( \lambda -m) \sinh k r}  \,.
   \end{eqnarray}
}\end{example}

The Darboux transformation for a generalized Coulomb
interaction is considered in \cite{DPSV-2002}.

 \section{Conclusion and outlook}

 In this paper only a small part of the
 properties of differential intertwining (Darboux) operators
 for the one-dimensional Dirac equation has been considered.
 We have shown that some properties known for the case of the
Schr\"odinger equation, like factorization of Hamiltonians, one-to-one
 correspondence between spaces of solutions of equations
 related with an intertwiner, supersymmetry, and determinant
 formulas for chains of transformations, also have their counterparts
  for the case of the Dirac equation. We have shown with numerous examples
  that this technique is as easily as in the case of the Schr\"odinger equation.
 Therefore, we hope that this paper will stimulate further
 investigations in this field.

To put an end to this paper, we would like to
 enumerate some subjects that, from our point of view, are worth to be investigated in
the future:
\begin{enumerate}
\item
To get purely scalar spherically symmetric
potentials.
\item
To find conditions for a chain of transformations to
produce regular potentials.
\item
To find interrelation between differential and integral
transformation operators.
\item
To find transformation of scattering data such as
 transmission and reflection coefficients, phase shifts.
\item
To consider transformations of periodical potentials.
\item
To apply this technique for describing the scattering
  of particles at high energies.
\end{enumerate}

 \section*{Acknowledgments}
This work has been partially
supported by the Spanish MCYT and the European FEDER (grant
BFM2002-03773), and also by Ministerio de Educaci\'on, Cultura y
Deporte of Spain (grant SAB2000-0240).


\begin{thebibliography}{00}

\bibitem{FernHussMeln}
 D.~J. Fern\'andez~C., V. Hussin, and B. Mielnik,
{\it Phys. Lett. A} {\bf 244} (1998),  309.


\bibitem{NegNiRos}
J. Negro, L.~M. Nieto, and O. Rosas-Ortiz,
{\it J. Math. Phys.} {\bf 41}  (2000), 7964.

\bibitem{FernNegNi}
 D.~J. Fern\'andez~C., J. Negro, and L.~M. Nieto,
{\it Phys. Lett. A} {\bf 275} (2000),  338.

\bibitem{Samsonov99}
B.~F. Samsonov, {\it Phys. Lett. A} {\bf 263} (1999),  274.

\bibitem{CRF}
 J.~C. Cari\~nena, A. Ramos, and D.~J.  Fern\'andez~C.,
{\it Ann.  Phys. NY} {\bf 292} (2001),  42.

\bibitem{FMRS}
D.~J. Fern\'andez~C., B. Mielnik, O Rosas-Ortiz, and B.~F. Samsonov,
{\it J. Phys. A} {\bf 35} (2002), 4279.

\bibitem{Andr}
 A.~A. Andrianov, N.~V. Borisov, M.~V.  Ioffe, and M.~I. Eides,
{\it Theor. Math. Phys.} {\bf 61} (1984), 17.

\bibitem{Wit}
E. Witten,
{\it Nucl. Phys. B} {\bf 185} (1981), 513;
{\it Nucl. Phys. B} {\bf 202} (1982), 253.

\bibitem{Darboux}
G. Darboux,
``Le\c cons sur la th\'eorie g\'en\'erale des surfaces et les
 application g\'eom\'etriques du calcul infinit\'esimale."
Deuxi\'eme partie. - Paris, Gautier-Villar et Fils.  1889;
 {\it Compt. Rend. Acad. Sci. Paris.} {\bf 94} (1882), 1343;
 {\it Compt. Rend. Acad. Sci. Paris.} {\bf 94} (1882), 1456.

\bibitem{MatveevSall}
V. Matveev and M. Salle,
 ``Darboux Transformations and Solitons",
New York, Springer, 1991.

\bibitem{Sukumar}
C. V. Sukumar,
{\it J. Phys. A} {\bf 18} (1985), 2937.


\bibitem{Pashnev}
V.~P. Berezovoy and A.~I. Pashnev,
{\it Theor. Math. Phys.} {\bf 74} (1988), 392.

\bibitem{Bay}
D. Baye, {\it Phys. Rev. Lett.} {\bf 58} (1987), 2738; L. U. Ancarani
and D. Baye, {\it Phys. Rev.} {\bf A46} (1992), 206; D. Baye, {\it Phys. Rev. A}
{\bf 48} (1993), 2040; D. Baye and J. -M. Sparenberg, {\it Phys. Rev.
Lett.} {\bf 73} (1994), 2789; G. L\'evai, D. Baye and J. -M.
Sparenberg, {\it J. Phys. A} {\bf 30} (1997), 8257;
J. -M. Sparenberg and D. Baye, {\it Phys. Rev. Lett.} {\bf 79}
(1997), 3802; H. Leeb, S. A. Sofianos, J. -M. Sparenberg and D. Baye,
{\it Phys. Rev. C} {\bf 62} (2000), 064003.

\bibitem{Schrodinger}
E. Schr\"odinger, {\it Proc. Roy. Irish. Acad. A.} {\bf 46} (1940), 9;
 {\it Proc. Roy. Irish. Acad. A.} {\bf 47} (1941), 53.

\bibitem{InfeldHall}
 H. Hull and T.~E. Infeld, {\it Phys. Rev.} {\bf 74} (1948), 905;
{\it Phys. Rev.} {\bf 59} (1941), 737;
{\it Rev. Mod. Phys.} {\bf 53} (1951), 21.

\bibitem{Anderson}A. Anderson,
{\it Phys. Rev. A} {\bf 43} (1991), 4602.

\bibitem{Stahlhofen}
A. A. Stahlhofen, {\it J. Phys. A} {\bf 27} (1994), 8279.

\bibitem{Sall-82}
M.~A. Salle, {\it Theor. Math. Phys.} {\bf 53} (1982), 1092.

\bibitem{Sall-87}
M.~A. Salle, {\it Zapiski Nauchnnih Seminarov LOMI} {\bf 161} (1987), 72.

\bibitem{DascalovChristov}V.~B. Daskalov and E.~Kh. Khristov,
{\it Inv. Probl.} {\bf 16} (2000), 247.

\bibitem{PratsToll} P. Prats and J.~S. Toll,
{\it Phys. Rev.} {\bf 113} (1959), 363.

\bibitem{LevitanSargsyan}
 M.~B. Levitan and I.~S. Sargsian,
  Introduction to Spectral Theory,
American Mathematical Siciety, Providence, Rhode Island, 1975.

\bibitem{SamsonovPecheritsin}
B.~F. Samsonov and A.A. Pecheritsin,
 {\it  Rus. Phys. J.} {\bf 43(1)} (2000), 48;
 {\it  Rus. Phys. J.} {\bf 45(1)} (2002), 14;
 {\it  Rus. Phys. J.} {\bf 45(1)} (2002), 74.


\bibitem{Crum}
M. M. Crum  {\it Quart. J. Math. Ser 2}, {\bf 6} (1955), 121.

\bibitem{BSTMF}
V.~G. Bagrov and B.~F. Samsonov,
{\it Theor. Math. Phys.} {\bf 104} (1995), 356.

\bibitem{Smirn}
V. Smirnov, ``Advanced Course of Mathematics", Vol. 5,
Phys.-Math. Publishing State House, Moscow, 1958.

\bibitem{RS}
M. Reed and B. Simon,
``Methods of Modern Mathematical Physics. I.
Functional Analysis", Academic, New York, 1972.

\bibitem{DS}
N. Dunford and J.~T. Schwartz,
``Linear Operators. Part II. Spectral Theory.
Self-Adjoint Operators in Hilbert Spaces",
Interscience, New York, 1963.

\bibitem{10} V.~G. Bagrov and B.~F. Samsonov,
{\it Phys. Part. Nucl.} {\bf 28} (1997), 951.

\bibitem{10a}
J. Beckers, N. Debergh and C. Gotti, {\it Helv. Phys. Acta} {\bf 71} (1998), 214.

\bibitem{pra-1998-v57-p93}
 Y. Nogami and F.~M. Toyama,
{\it Phys.~Rev. A} {\bf 57} (1998), 93.

\bibitem{Bagchi}
B.~K. Bagchi, ``Supersymmetry in Quantum and Classical Mechanics",
 Chapman and Hall, New York, 2001.

\bibitem{Nogami-93}
Y. Nogami and F.~M. Toyama, {\it Phys. Rev. A} {\bf 47} (1993), 1708.

\bibitem{NogToy98}
F.~M. Toyama and Y. Nogami,
 {\it Phys. Rev. A} {\bf 57} (1998), 93.

\bibitem{Thaler}
B. Thaler, ``The Dirac Equation", Springer, Berlin, 1992.

\bibitem{Toyama-97}
Y. Nogami and F.~M. Toyama,
 {\it J. Phys. A} {\bf 30} (1997), 2585.

\bibitem{Benvegnu}
 S. Benvegnu,
 {\it J. Math. Phys.} {\bf 38} (1997), 556.

 \bibitem{Dominguez} F. Dom\'{\i}nguez-Adame,
 {\it Am. J. Phys.} {\bf 58} (1990), 886.

 \bibitem{Khalilov}
 Ho Choon-Lin and V.~R. Khalilov,
 {\it Phys. Rev. D} {\bf 63} (2000), 027701.

 \bibitem{Weidman}
 J. Weidman,
 ``Spectral Theory of Ordinary Differential Operators",
{\it Lect. Notes Math., Vol. 1258}, Springer, Berlin, 1987.

 \bibitem{DPSV-2002}
 N. Debergh, A. A.  Pecheritsin,
B. F. Samsonov and B. Van Den Bossche,
 {\it J.~Phys.~A} {\bf 35} (2002), 3279.

 \end{thebibliography}
\end{document}